\documentclass[aps,nofootinbib,showpacs,floats,letterpaper,floatfix,groupedaddress,eqsecnum]{revtex4}

\usepackage{bm}
\usepackage{amsfonts}
\usepackage{latexsym}
\usepackage[latin1]{inputenc}
\usepackage{graphicx}
\usepackage{amsmath}
\usepackage{epsfig}
\usepackage{amsbsy}

\newcommand{\rhz}{\rho_{0}}
\newcommand{\td}{\text{d}}
\newcommand{\oda}[2]{\frac{\td #1}{\td #2}}

\def\jnl@style{\it}
\def\aaref@jnl#1{{\jnl@style#1}}

\def\aaref@jnl#1{{\jnl@style#1}}

\def\aj{\aaref@jnl{AJ}}                   
\def\apj{\aaref@jnl{ApJ}}                 
\def\apjl{\aaref@jnl{ApJ}}                
\def\apjs{\aaref@jnl{ApJS}}               
\def\apss{\aaref@jnl{Ap\&SS}}             
\def\aap{\aaref@jnl{A\&A}}                
\def\aapr{\aaref@jnl{A\&A~Rev.}}          
\def\aaps{\aaref@jnl{A\&AS}}              
\def\mnras{\aaref@jnl{MNRAS}}             
\def\prd{\aaref@jnl{Phys.~Rev.~D}}        
\def\prl{\aaref@jnl{Phys.~Rev.~Lett.}}    
\def\qjras{\aaref@jnl{QJRAS}}             
\def\skytel{\aaref@jnl{S\&T}}             
\def\ssr{\aaref@jnl{Space~Sci.~Rev.}}     
\def\zap{\aaref@jnl{ZAp}}                 
\def\nat{\aaref@jnl{Nature}}              
\def\aplett{\aaref@jnl{Astrophys.~Lett.}} 
\def\apspr{\aaref@jnl{Astrophys.~Space~Phys.~Res.}} 
\def\physrep{\aaref@jnl{Phys.~Rep.}}      
\def\physscr{\aaref@jnl{Phys.~Scr}}       

\begin{document}


\title[Relativistic Ghost Fluid method]{Numerical simulations of
  interfaces in relativistic hydrodynamics}


\author{S.~T. Millmore}
\affiliation{Department of Mathematics, University of Southampton, Southampton SO17 1BJ, UK}

\author{I. Hawke}
\affiliation{Department of Mathematics, University of Southampton, Southampton SO17 1BJ, UK}


\date{\today}


\begin{abstract}

  We consider models of relativistic matter containing sharp
  interfaces across which the matter model changes. These models will
  be relevant for neutron stars with crusts, phase transitions, or for
  viscous boundaries where the length scale is too short to be
  modelled smoothly. In particular we look at numerical techniques
  that allow us to evolve stable interfaces, for the interfaces to
  merge, and for strong waves and shocks to interact with the
  interfaces. We test these techniques for ideal hydrodynamics in
  special and general relativity for simple equations of state,
  finding that simple level set-based methods extend well to
  relativistic hydrodynamics.

\end{abstract}


\pacs{
04.25.D-, 
04.40.-b 
}

\maketitle


\section{Introduction} \label{sec:Intro}


Numerical simulations of compact objects such as neutron stars (NSs)
are widely used to predict the nonlinear dynamics of, for example,
binary merger or collapse \cite{lrr-2008-7}. Accurate and detailed
simulations are thought necessary for constructing gravitational wave
templates for use in detectors such as LIGO and VIRGO
\cite{LIGOReview,Acernese2006}. The dynamics and wavesignal are in
large part determined by the structure of the NS, which in turn is
determined by the bulk equation of state (EOS) and the detailed
microphysics in key, narrow, regions near the surface, crust and
internal interfaces.

Most current models of NSs used in numerical simulations use single
ideal perfect fluids (plus magnetic fields). When extending this
model, the important physics may be different in specific regions of
the NS (the surface and exterior, the crust and the core, regions of
superconductivity, or local phase transitions). Some effects, such as
viscosity, will play a role only in localised regions. In all of these
situations there will be complex behaviour concentrated in a thin
interface layer. The question is how these interfaces can be modelled
accurately and practically in a numerical simulation.

Whether an interface can be simulated practically depends on the
length scales involved, both in the underlying physics and in the
simulation. In single fluid stars the surface (interface between fluid
and vacuum) is genuinely sharp. Even in more realistic situations the
physical length scales may be extremely short. A more quantitative
example is given by the Ekman layer in hot NS cores which is relevant
for proto NS modelling. In this case the length scale has been
estimated by \citep{Andersson:2004aa} to have the form $\delta \sim
[\eta / ( \rho \Omega) ]^{1/2}$ with the viscosity coefficient $\eta$
being highly temperature dependent, but with $\eta \sim 10^{13} -
10^{19} \text{ g cm}^{-1}\text{ s}^{-1}$ being reasonable for $T >
10^9 \text{ K}$.  Working in cgs units and choosing $\eta \approx
10^{16} \text{ g cm}^{-1}\text{ s}^{-1}$ as a representative value
leads to $\delta \sim 10 \Omega^{-1/2} \text{ cm}$ for $\rho \sim
10^{14} \text{ g cm}^{-3}$.

We then need to consider what physical length scales can be resolved
in feasible numerical simulations. Focusing on finite difference
simulations using current technology, we assume that it is practical
to simulate for $N \sim 10^{10}$ timesteps and that we wish to cover
$10 \text{ ms}$ in physical time. Then our finest timestep must be
$\sim 10^{-12} \text{ s}$ which means, by the CFL criteria, that are
smallest grid spacing must be $\sim 10^{-1} \text{ cm}$. Standard
numerical methods smear interfaces and contact discontinuities with
time (\citep{harten1978}) with the width of a sharp interface being
spread over $N^{1/(p+1)}$ zones after $N$ timesteps, where $p$ is the
order of accuracy of the method. Making the optimistic choice of $p=4$
we see that the interface is smeared over $10^2$ zones with a physical
width of $\sim 10 \text{ cm}$. Even with these optimistic assumptions
the numerical simulation will be unable to distinguish numerical and
physical effects for NSs with even moderate rotation rates.

This argument suggests that it is impractical and inefficient to model
the detailed physics involved in these interface regions
directly. However, we still need to incorporate these effects as best
we can within our numerical simulations. To do this we will model
these interfaces as being genuinely sharp; that is, they have zero
width. The additional physics will then be incorporated through the
dynamical behaviour of the interfaces, and the boundary conditions
imposed there.

In this paper we will use a number of single ideal fluid components
separated by sharp material interfaces. We will work in full general
relativity using ideal hydrodynamics but restrict to $1+1$ dimensions
for simplicity. The model and a way to simulate it numerically is
outlined in section~\ref{sec:model}.  A specific numerical
implementation is described in section~\ref{sec:implement}. Finally we
show a variety of tests illustrating the advantages and limitations of
this simple implementation in section~\ref{sec:tests}.

Throughout the paper we use a signature of $(-,+,+,+)$ and work in
geometric ($G = c = M_{\odot} = 1$) units.


\section{Modelling sharp interfaces}
\label{sec:model}


\subsection{The continuum model}
\label{sec:model_cont}

The continuum model that we wish to simulate considers the spacetime
as being composed of separate regions denoted $\Omega_X$. On each
region the physical matter model (as described by a stress-energy
tensor and equations of motion) may be different. Each region obeys
the Einstein equations and any additional equations of motion required
for the matter. In the present paper we restrict ourselves to vacuum
or to single ideal relativistic perfect fluids with the stress-energy
tensor
\begin{equation}
  \label{eq:Tmunu_fluid}
  T^{\mu \nu} = 
  \rho_0 h u^{\mu} u^{\nu} + p g^{\mu \nu},
\end{equation}
where $\rho_0$ is the rest-mass density, $h$ is the specific enthalpy
defined by
\begin{equation}
  \label{eq:Tmunu_h}
  h = \left( 1 + \varepsilon + \frac{p}{\rho_0} \right),
\end{equation}
with $\varepsilon$ the specific internal energy, $p$ the pressure, and
$u^{\mu}$ the four-velocity.

The full spacetime dynamics is described by the union of all the
regions $\Omega_X$. At and across the boundaries $\partial \Omega_X$
the Einstein equations must be satisfied, and from the results
of~\citep{Barnes:2004if} we note that this should not introduce
discontinuities to invariant spacetime quantities, and hence we expect
no additional problems from the spacetime. Therefore all that remains
to describe the model is to give the conditions satisfied by the
matter model at the boundaries $\partial \Omega_X$, and the conditions
specifying the location of the boundaries.  In what follows we shall
use a standard $3+1$ splitting approach, so the spacetime is foliated
into slices. The spacetime regions $\Omega_X$ and their boundaries
become spacelike volumes within the slice.

For this paper we consider the simplest possible situation for the
boundary and interface conditions. We will assume that there is force
balance at the boundary and no diffusion through it. We will assume
that there is no surface tension or other non-ideal effect acting on
or within the boundary. This means that the boundary will advect with
the neighbouring fluid velocities, and the boundary condition will
depend on the pressure normal to it.

\subsection{Numerical model}
\label{sec:model_num}

As noted above we only need to specify the boundary location and the
conditions on the matter model at and across the boundary.  We will
first discuss the conditions applied to the location of the boundary.

\subsubsection{Boundary location}
\label{sec:model_num_boundary_locn}

In order to impose a boundary condition at $\partial \Omega_X$ we
first need to know where the boundary is. We assume that the location
is known in the initial slice. We need to find the best method of
identifying the location of the boundary at any later time. We note
that this method must be able to deal with complex changes in topology
of any interface, particularly in merger simulations.

The standard technique for dealing with sharp features such as shocks
and interfaces in a single perfect fluid is to \emph{capture} them. In
this approach the interface is smeared over a (hopefully small) number
of points, with the appropriate solution found in the continuum
limit. This is incompatible with the approach we are taking here. As
well as the length scale issues discussed in the introduction, there
are serious numerical problems when the model changes at the interface
and the physics of the mixture are not taken into account, as
illustrated in appendix~\ref{sec:appendix_colourfn}.

The alternative is to explicitly track or locate the interface. This
is a standard problem in numerical relativity, as event and apparent
horizons must be located in numerical simulations (for a review
see~\citep{Thornburg:2006zb}). Of particular relevance to us are the
techniques used for event horizons (see for
example~\citep{Diener:2003jc}) where the interface is implicitly
tracked and complex changes of topology can be dealt with. The
techniques used borrow substantially from those in Newtonian CFD which
are precisely used to deal with the problem we are interested in here
(see~\citep{OsherFedkiw2003,Sethian1996} for detailed descriptions).

Explicitly, we will use a \emph{level set} to track the boundary
$\partial \Omega_X$. That is, we introduce a scalar function $\phi$
defined over the entire spacetime. The location of the boundary is
implicitly given by those points where $\phi$ vanishes, i.e.\ by
\begin{equation}
  \label{eq:levelset_boundary}
  \partial \Omega_X = \left\{ {\bf x}: \phi({\bf x}) = 0 \right\}.
\end{equation}
The condition defining the location of the boundary $\partial
\Omega_X$ then becomes (after the $3+1$ split) an evolution equation
for the scalar function $\phi$. This evolution equation is arbitrary
\emph{except} at the boundary. The equation can therefore be chosen
for modelling simplicity or numerical convenience, provided that it
reproduces the correct behaviour at the boundary.

In this paper we consider the boundary as being advected with
the fluid, i.e.\ it will be Lie dragged by the 4-velocity
$u^{\mu}$. The obvious evolution equation for the level set is
\begin{equation}
  \label{eq:levelset_eom1}
  {\cal L}_{\bf u} \phi = 0.
\end{equation}
However, the evolution of the level set can then lead to complex
behaviour for $\phi$ away from the boundary, as noted by
e.g.~\citep{Diener:2003jc,OsherFedkiw2003}. This complexity is
irrelevant to the physical behaviour. Methods for ``renormalizing''
the level set (which changes the behaviour of $\phi$ away from the
boundary) are typically used to avoid problems. In $1+1$ dimensions,
and when only a single boundary is present, these problems can be
avoided by using a constant vector instead of the fluid
4-velocity. Obviously this constant vector must be chosen to match the
velocity at the boundary, i.e.\ equation~(\ref{eq:levelset_eom1})
becomes
\begin{equation}
  \label{eq:levelset_eom2}
  {\cal L}_{\left. \bf u \right|_{\partial \Omega_X}} \phi = 0.
\end{equation}

\subsubsection{Boundary condition}
\label{sec:model_num_boundary_condn}

To complete the description of the model we need to describe the
conditions imposed at the boundary. There are a number of properties
that we would like these conditions to satisfy. We need the conditions
to be compatible with the continuum model. We would like the
conditions to approximate, as closely as possible, the additional
microphysics. We need the boundary conditions to be practical in a
numerical simulation; in particular, it must be possible to use them
without introducing unphysical oscillations due to any discontinuities
present. We would like the boundary conditions to be as simple as
possible.

The condition that we study in this paper is the simplest successful
condition proposed in Newtonian CFD; the \emph{Ghost Fluid} condition
proposed by~\citep{Fedkiw1999457}. It is intended to model the simple
situation described above; two ideal fluids separated by an interface
with no diffusion. It is simple and practical to implement, but is not
correct in all circumstances; studies such
as~\citep{Liu2003651,Liu2005193} have shown that incorrect shock
structures can be computed when strong shocks hit the interface
between extremely different EOSs. We use the Ghost Fluid boundary
condition to illustrate the general approach.

The Ghost Fluid approach considers the fluid on each spacetime region
$\Omega_X$ separately. To impose boundary conditions at $\partial
\Omega_X$ the fluid describing the matter within the spacetime region
$\Omega_X$ is artificially extended through the interface into the
neighbouring region(s). This is analogous to the imposition of
boundary conditions using ghost zones or ghost points. The boundary
conditions are then imposed by providing values for the artificially
extended fluid. In this approach a condition is given on the pressure,
the velocity and the entropy, from which all other quantities can be
recovered (given an EOS).

\begin{figure}[htbp]
  \centering
  \includegraphics[width=0.7\textwidth]{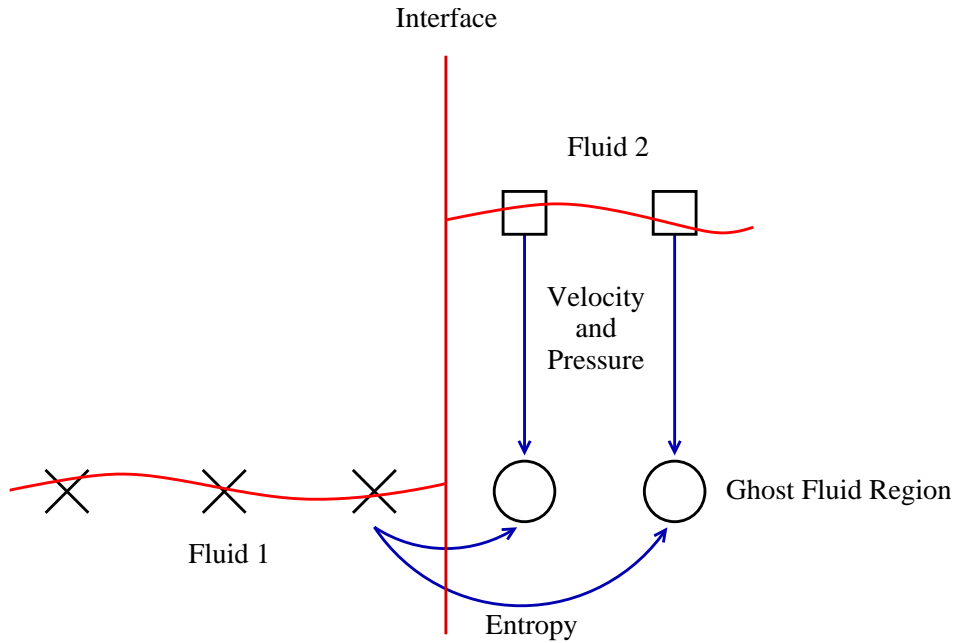}
  \caption{Illustrating the Ghost Fluid boundary condition in one
    dimension. The boundary condition is to be applied to fluid
    1. This is done by creating an artificial ``ghost'' fluid in the
    region covered by fluid 2. This ghost fluid obeys the same
    equations of motion and EOS as fluid 1 and is used to implicitly
    capture the boundary conditions at the interface. The pressure and
    (normal) velocity are copied from fluid 2. The entropy (and, in
    higher dimensions, the tangential velocity) are extrapolated from
    fluid 1. The same procedure is applied to fluid 2.}
  \label{fig:ghostfluid1}
\end{figure}

The precise conditions imposed at the interface are the continuity of
the pressure and the normal velocity, and that the flow is
isentropic. In principal for a material interface we would expect the
pressure and normal component of the velocity to be continuous;
entropy and the tangential velocity may jump. To follow this as
closely as possible the pressure and normal velocity component are
copied from the fluid in the \emph{neighbouring} region $\Omega_Y$,
whilst the entropy and tangential velocity are extrapolated. This is
illustrated in $1+1$ dimensions in figure~\ref{fig:ghostfluid1}.


\section{Implementation} 
\label{sec:implement}


\subsection{Formulation}
\label{sec:implement_form}

\subsubsection{Hydrodynamics - flat space}
\label{sec:implement_form_hydro_flat}

In order to best compare to standard relativistic hydrodynamics
simulations we use the conserved formulation of~\cite{Banyuls97}. In
$1+1$ Minkowski spacetime using Cartesian coordinates and the
test-fluid approximation this is written as
\begin{equation}
  \label{eq:form_valencia_flat}
  \partial_t {\bf q} + \partial_x {\bf f}({\bf q}) = {\bf 0},
\end{equation}
where the conserved variables ${\bf q}$ and fluxes ${\bf f}$ are given
in terms of the primitive variables introduced in
section~\ref{sec:model} by
\begin{subequations}
  \label{eq:form_valencia_variables}
  \begin{align}
    \label{eq:form_valencia_cons_vars}
    {\bf q} & = \left( D, S_x, \tau \right)^T
     = \left( \rho_0 W, \rho_0 h W^2 v_x, \rho_0 h W^2 - p - \rho_0 W
    \right)^T, \\
    \label{eq:form_valencia_fluxes}
    {\bf f} & = \left( D v^x, S_x v^x + p, \left( \tau + p \right) v^x
    \right)^T,
  \end{align}
\end{subequations}
where we have introduced the 3-velocity
\begin{equation}
  \label{eq:form_valencia_vel}
  v^i = \frac{1}{\alpha} \left( \frac{u^i}{u^t} + \beta^i \right) =
  \frac{u^i}{u^t},
\end{equation}
where the shift $\beta^i$ and lapse $\alpha$ resulting from the $3+1$
split take the trivial values in this case. We have also introduced
the Lorentz factor
\begin{equation}
  \label{eq:form_valencia_W}
  W = \left( 1 - v_i v^i \right)^{-1/2}.
\end{equation}

To close the system an equation of state is required. In this paper we
use the simple $\gamma$-law equation of state
\begin{equation}
  \label{eq:EOS}
  p( \rho_0, \epsilon ) = (\gamma - 1) \rho_0 \epsilon,
\end{equation}
where $\gamma$ is the ratio of specific heats. Different regions of
spacetime may have different equations of state which is modelled
simply by providing different values of $\gamma$.

The level set equation~\eqref{eq:levelset_eom1} becomes
\begin{equation}
  \label{eq:form_valencia_levelset}
  \partial_t \phi + v^x \partial_x \phi = 0.
\end{equation}
This equation is not in conservation law form. Whilst it can be
written in conservation law form this would have no advantage, as it
is only the zeros of the level set function that have meaning. The
non-conservative form of equation~(\ref{eq:form_valencia_levelset}) is
evolved directly.

\subsubsection{Hydrodynamics - spherical symmetry}
\label{sec:implement_form_hydro_sph}

In $1+1$ spherical symmetry we follow
e.g.~\cite{Romero:1995cn,NeilsenThesis,NobleThesis} in using
polar-areal coordinates. The line element is written
\begin{equation}
  \label{eq:form_valencia_sph_line_element}
  \td s^2 = - \alpha^2(t, r) \td t^2 + a^2(t, r) \td r^2 + r^2 \left(
    \td \theta^2 + \sin^2 \theta \, \td \phi^2 \right). 
\end{equation}
The 3-velocity becomes
\begin{equation}
  \label{eq:form_valencia_vel_sph}
  v^r = \frac{u^r}{\alpha u^t}.
\end{equation}
The conservation law form is written
\begin{equation}
  \label{eq:form_valencia_sph}
  \partial_t \left( a {\bf q} \right) + \frac{1}{r^2} \partial_r
  \left( \alpha a r^2 {\bf f}({\bf q}) \right) = {\bf s}({\bf q}),
\end{equation}
where the variables, fluxes and sources are
\begin{subequations}
  \label{eq:form_valencia_variables_sph}
  \begin{align}
    \label{eq:form_valencia_cons_vars_sph}
    {\bf q} & = \left( D, S_r, \tau \right)^T = \left( \rho_0 W, \rho_0 h
      W^2 v_r, \rho_0 h W^2 - p - \rho_0 W \right)^T, \\
    \label{eq:form_valencia_fluxes_sph}
    {\bf f} & = \left( D v^r, S_r v^r + p, \left( \tau + p \right) v^r
    \right)^T, \\
    \label{eq:form_valencia_sources_sph}
    {\bf s} & = \alpha a \left( 0, -\frac{a^2 m}{r^2} \left( S_r v^r +
        \tau + p + D \right) + \frac{2 p}{r}, -\frac{m}{r^2} S_r
    \right)^T,
  \end{align}
\end{subequations}
where the spacetime quantity $m = \frac{r}{2} \left( 1 - a^{-2}
\right)$ is the mass aspect function. The spacetime quantities satisfy
the equations
\begin{subequations}
  \label{eq:form_valencia_spacetime}
  \begin{align}
    \label{eq:form_valencia_spacetime_a_evol}
    \partial_t a & = - 4 \pi r \alpha a S_r, \\
    \label{eq:form_valencia_spacetime_a_constr}
    \partial_r a & = a^3 \left[ 4 \pi r \left( \tau + D \right) -
      \frac{m}{r^2} \right], \\
    \label{eq:form_valencia_spacetime_alpha_constr}
    \partial_r \log(\alpha) & = a^2 \left[ 4 \pi r \left( S_r v^r + p
      \right) + \frac{m}{r^2} \right].
  \end{align}
\end{subequations}
We will evolve $a$ using
equation~(\ref{eq:form_valencia_spacetime_a_evol}) and solve for the
lapse using the slicing
condition~(\ref{eq:form_valencia_spacetime_alpha_constr}), whilst the
error in the Hamiltonian
constraint~(\ref{eq:form_valencia_spacetime_a_constr}) will be used to
check the accuracy of the simulations. 

To reduce problems near the coordinate singularity at the origin the
equations are rewritten as
\begin{equation}
  \label{eq:form_valencia_sph_rewrite}
  \partial_t \left( a {\bf q} \right) + \frac{1}{3} \partial_{r^3}
  \left( \alpha a r^2 {\bf f}^{(1)}({\bf q}) \right) + \partial_r
  \left( \alpha a {\bf f}^{(2)} ({\bf q}) \right) = \bar{\bf s}({\bf q}),
\end{equation}
where the revised fluxes and sources are
\begin{subequations}
  \label{eq:form_valencia_sph_rewrite_flux_source}
  \begin{align}
    \label{eq:form_valencia_sph_rewrite_flux1}
    {\bf f}^{(1)} & = \left( D v^r, S_r v^r, \left( \tau + p \right)
      v^r \right)^T, \\
    \label{eq:form_valencia_sph_rewrite_flux2}
    {\bf f}^{(2)} & = \left( 0, p, 0 \right)^T, \\
    \label{eq:form_valencia_sph_rewrite_source}
    \bar{\bf s} & = \alpha a \left( 0, -\frac{a^2 m}{r^2} \left( S_r
        v^r + \tau + p + D \right), -\frac{m}{r^2} S_r \right)^T.
  \end{align}
\end{subequations}

The level set equation~\eqref{eq:levelset_eom1} becomes
\begin{equation}
  \label{eq:form_valencia_levelset_sph}
  \partial_t \phi + \alpha v^r \partial_r \phi = 0.
\end{equation}
As in flat space, this non-conservative form is evolved directly. 

\subsection{Numerical methods}
\label{sec:implement_numerics}

To evolve in time the method of lines is used with a second order
Runge-Kutta algorithm.

\subsubsection{Hydrodynamics}
\label{sec:implement_numerics_hydro}

To illustrate the utility of the approach and the Ghost Fluid boundary
condition we have used standard High Resolution Shock Capturing (HRSC)
methods to solve the hydrodynamics
equations~(\ref{eq:form_valencia_flat}) and
(\ref{eq:form_valencia_sph_rewrite}). The domain is split into cells,
and the conserved variables updated using
\begin{equation}
  \label{eq:cons_update}
  \oda{}{t} \hat{\bf q}_i = \frac{1}{\Delta x_i} \left[ \hat{\bf
      f}_{i-1/2} - \hat{\bf f}_{i+1/2} \right].
\end{equation}
Following e.g.~\cite{Hawke:2005zw} we concentrate on
Reconstruction-Evolution methods, where the variables are
reconstructed to cell boundaries and a Riemann problem (approximately)
solved to find the inter-cell flux $\hat{\bf f}_{i \pm 1/2}$. For
reconstruction we use slope-limited TVD methods with van Leer's
monotized-centred (MC) limiter (\cite{vanLeer73}), or the PPM
method~(\cite{Colella1984174,Marti19961}). For simplicity we use the
HLLE flux formula~\cite{einfeldt:294} in what follows, although we
have seen no qualitative difference when the Marquina flux
formula~\cite{Donat199642} is used instead.

\subsubsection{Level set}
\label{sec:implement_numerics_level_set}

The evolution of the level set
(equation~\eqref{eq:form_valencia_levelset} or
\eqref{eq:form_valencia_levelset_sph}) cannot use the same numerical
methods as the equation is not in conservation law form. Instead we
shall follow the methods outlined in~\cite{OsherFedkiw2003}, evolving
the level set equation as a Hamilton-Jacobi equation. General
high-order methods for evolving Hamilton-Jacobi equations can be
found, for example, in~\cite{Shu99}.

The general method for Hamilton-Jacobi equations of the form
\begin{equation}
  \label{eq:hj_form}
  \partial_t \phi + H(\partial_x \phi) = 0
\end{equation}
that works even when the level set $\phi$ is not differentiable is to solve
\begin{equation}
  \label{eq:hj_form_numerical}
  \oda{}{t} \phi_i + \hat{H} \left( D^x_{+} \phi_i, D^x_{-} \phi_i
  \right) = 0,
\end{equation}
where $\phi_i$ is the numerical value of $\phi$ at gridpoint $x_i$,
$\hat{H}$ the numerical Hamiltonian, assumed Lipschitz continuous, and
$D^x_{\pm}$ appropriate winded approximations of $\partial_x \phi$.

It is simple to use ENO or WENO methods to find the appropriate
$D^x_{\pm}$ operators (explicit expressions for low order are given
by~\cite{OsherFedkiw2003}). In the $1+1$ examples studied here it has
been sufficient to use first order accurate upwind differencing. We
would not expect this to be sufficient in more complex examples or in
higher dimensions.

We have also used the simple Lax-Friedrichs approximation to the
Hamiltonian,
\begin{equation}
  \label{eq:hj_numerical_lf}
  \hat{H}(u_{+}, u_{-}) = H \left( \frac{u_{+} + u_{-}}{2} \right) -
  \hat{\alpha} \left( \frac{u_{+} - u_{-}}{2} \right),
\end{equation}
where $\hat{\alpha}$ is the maximum absolute value of $H$ over the
entire domain. In higher dimensions it may be costly to compute
$\hat{\alpha}$ over the whole domain, so a \emph{local} Lax-Friedrichs
method (where the maximum value of $\hat{\alpha}$ over the numerical
stencil is used) can be used instead.

\subsubsection{Ghost Fluid boundary condition}
\label{sec:implement_ghostfluid}

Figure~\ref{fig:ghostfluid1} has already been given to illustrate how
the Ghost Fluid boundary condition is implemented. To be explicit, we
consider a numerical grid $\left\{ x_i \right\}$ on which the level
set $\phi_i$ changes sign between points $I, I+1$. We apply the Ghost
Fluid boundary condition to fluid 1, which covers the domain to the
left of the interface, using where necessary information from fluid 2,
which covers the domain to the right of the interface. We will apply
the boundary condition to \emph{at least} $p+1$ ``ghost'' points
extending from $x_{I+1}$ to $x_{I+p+1}$, where $p$ is the stencil size
of the numerical method. The reason for the use of $p+1$ ghost points
is that at least one additional point is needed if the interface
should move one point to the right during the evolution step.

The normal component of the velocity and the pressure of fluid 1,
$v^{(1)}$ and $p^{(1)}$ respectively, are copied from fluid 2.  The
entropy is extrapolated from fluid 1. In the Ghost Fluid method zero
order extrapolation normal to the interface is used. In the $1+1$
dimensional cases here, the extrapolation is a simple copy. Explicitly
we have
\begin{subequations}
  \label{eq:ghostfluid_vp}
  \begin{align}
    \label{eq:ghostfluid_v}
    v^{(1)}_j & = v^{(2)}_j, & j & = I+1, \dots, I+p+1, \\
    \label{eq:ghostfluid_p}
    p^{(1)}_j & = p^{(2)}_j, & j & = I+1, \dots, I+p+1, \\
    \label{eq:ghostfluid_s}
    s^{(1)}_j & = s^{(1)}_I, & j & = I+1, \dots, I+p+1.
  \end{align}
\end{subequations}
All other hydrodynamical quantities are found from the pressure,
entropy, and the equation of state.

\subsubsection{The evolution step}
\label{sec:implement_numerics_step}

In simple cases such as flat space or where free evolution of
spacetime quantities is used, the evolution step is now simply a
matter of updating both the fluid and level set. However, should key
spacetime quantities (such as $a$ for the case given in
section~\ref{sec:implement_form_hydro_sph}) be updated using non-local
equations (for example, the
constraint~(\ref{eq:form_valencia_spacetime_a_constr}) could be used
to give $a$) then the order of the update becomes important. For this
reason we ensure that we update the level set using the methods of
section~\ref{sec:implement_numerics_level_set} before the
hydrodynamical quantities can be updated using the methods of
section~\ref{sec:implement_numerics_hydro}.

We represent a complete evolution step, assuming a first order
time-stepping algorithm, as:
\begin{enumerate}
\item Given data $\left\{ \hat{\bf q}^n_i \right\}$ for the
  hydrodynamical variables and  $\left\{ \phi^n_i \right\}$ for the
  level set and timestep $n$.
\item Find interface locations $\left\{ x_j^n : \phi = 0 \right\}, j =
  0, \dots, N^n$, where $N^n-1$ is the number of interfaces at time
  $t^n$. For simplicity we denote the boundaries of the computational
  domain by $x_0^n$ and $x_{N^n}^n$ respectively.
\item If there is only one genuine interface approximate the velocity
  $v$ there.
\item Compute the updated level set $\phi^{n+1}$, either using the
  interface velocity approximated in the previous step, or using the
  velocity of the fluid at each point.
\item Compute the new interface locations $\left\{ x_j^{n+1} : \phi =
    0 \right\}, j = 0, \dots, N^{n+1}$.
\item For each domain $\Omega_j^n = [x^n_j, x^n_{j+1}], j = 0, \dots,
  N^n-1$:
  \begin{enumerate}
  \item Check the domain still exists at the next time step,
    $\Omega_j^{n+1} \ne \left\{ \emptyset \right\}$. If it is, go to
    the next domain.
  \item Apply the Ghost Fluid boundary condition to both boundaries,
    extending the domain by a sufficient number of points, as given in
    section~\ref{sec:implement_ghostfluid}.
  \item Compute the inter-cell fluxes on this extended domain.
  \item Restrict the updated variables to the domain
    $\Omega_j^{n+1} = [x^{n+1}_j, x^{n+1}_{j+1}]$.
  \item Compute primitive variables and apply physical or symmetry
    boundary conditions as required.
  \end{enumerate}
\end{enumerate}
Extending this to higher order in time is trivial with method of lines
type methods.


\section{Tests} 
\label{sec:tests}


The tests used here are intended to see whether the Ghost Fluid method
works in relativistic situations as it does in Newtonian theory. This
will indicate whether this is a viable approach for numerical
simulations of more complex models of NS interactions such as
mergers. The simplest test is a check that the method would keep a
single non-interacting material interface stable. This is provided in
appendix~\ref{sec:appendix_colourfn} as part of an illustration of why
capturing an interface is not straightforward numerically, as
discussed in section~\ref{sec:model_num_boundary_locn}.

We start by looking at the interaction between strong nonlinear
features and interfaces in flat space, using the model of
section~\ref{sec:implement_form_hydro_flat}. These ensure that no
spurious oscillations are introduced and see how accurately the method
captures transient and long-lived smooth features. We then move on to
full GR tests, where we can assess the accuracy and effectiveness with
simple spherically symmetric NS models, both stable and nonlinearly
perturbed.

\subsection{Flat space tests}
\label{sec:tests_flat}

\subsubsection{Shock-interface interaction}
\label{sec:tests_flat_shock}

Two tests are considered. In both the domain is $x \in [0, 1]$. In the
first a stable material interface is set up in the middle of the
domain across which the equation of state changes due to a change in
$\gamma$. A mild shock is injected from the left edge of the
domain. By time $t = 1$ the shock has interacted with the interface,
being reflected and transmitted. In this test, which is a modified
relativistic analogue of test B of~\cite{Fedkiw1999457}, the precise
initial data is
\begin{equation}
  \label{eq:tests_flat_shock_ID}
  \left\{ \begin{array}{clclclclc}
    \rhz = &  1.3346, \quad & v = & 0.1837, \quad & p = & 1.5, \quad
    & \gamma = & 1.4 \qquad & x < 0.05 \\
    \rhz = &  1, \quad  & v = & 0, \quad & p = & 1, \quad
    & \gamma = & 1.4 \qquad & 0.05 \le x < 0.5 \\
    \rhz = &  1, \quad  & v = & 0, \quad & p = & 1, \quad
    & \gamma = & 1.67 \qquad & x \ge 0.5.
  \end{array} \right. 
\end{equation}

\begin{figure}[htbp]
  \centering
  \includegraphics[width = 0.7\textwidth]{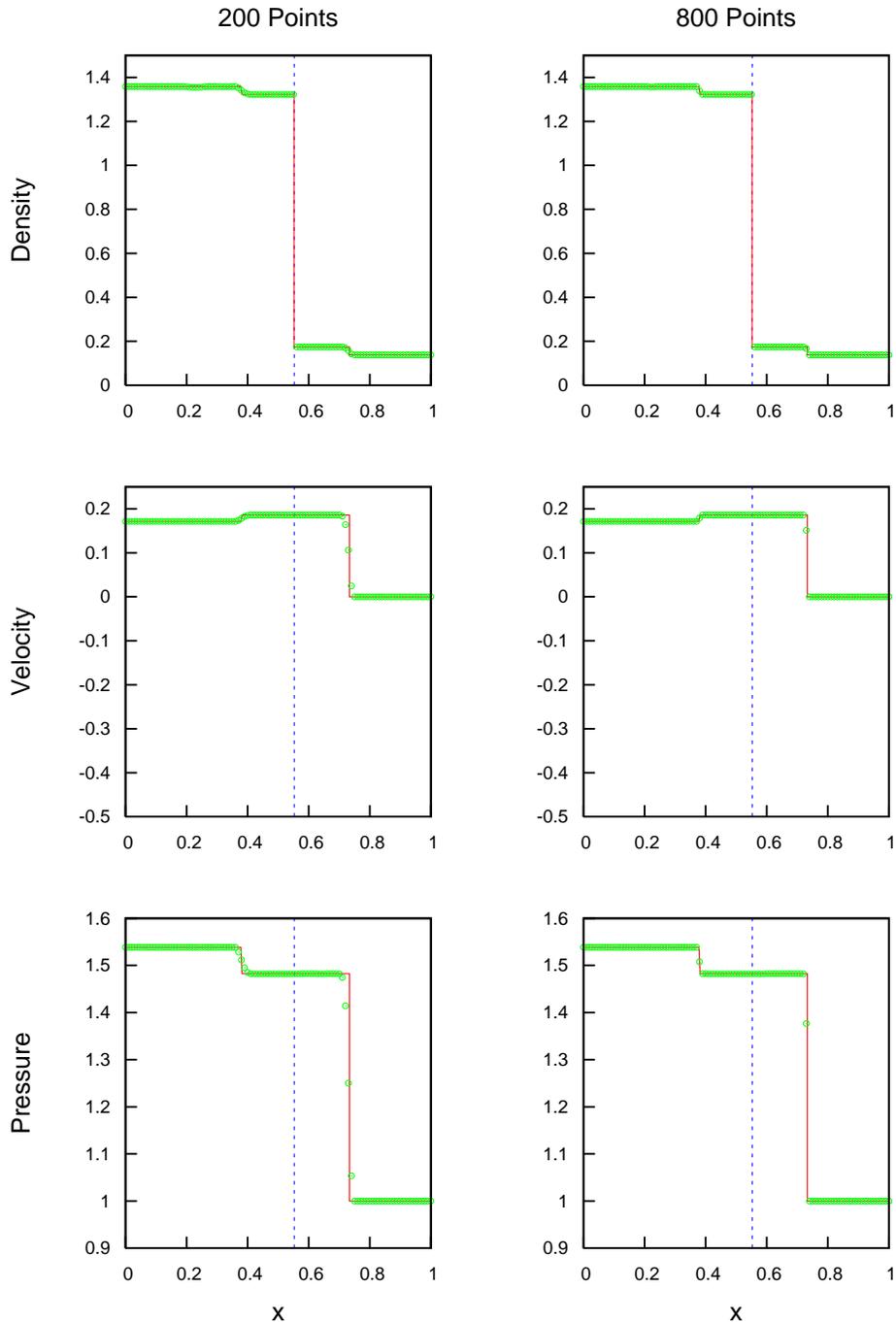}  
  \caption{Results from an isolated shock hitting a contact
    discontinuity, (`test B' from \cite{Fedkiw1999457}).  The
    transmitted shock is very obvious in the velocity and pressure
    profiles, and convergence of the solution at this point can be
    seen.  The reflected rarefaction wave is considerably smaller, but
    can still be seen.  The contact discontinuity, shown by the dashed
    line, is captured correctly. TVD reconstruction using the MC
    limiter is used, and only 100 points are shown in each plot.}
  \label{fig:tests_flat_shock}
\end{figure}
Results from this test are shown in
figure~\ref{fig:tests_flat_shock}. For simplicity we use TVD
reconstruction with the HLLE Riemann solver.  The transmitted shock
and reflected rarefaction are clearly seen, with the results
converging towards the exact solution. The material interface has
propagated to the right due to the interaction, and this is also
captured accurately. No spurious oscillations have been generated
either at the shock or at the material interface.

A considerably more difficult test that does not have an exact
solution was suggested by~\cite{Wang2004528}. This has a ``slab'' of
helium/air mixture initially at rest in air. A moderate shock then
strikes the slab, with the resulting shocks and rarefaction
interacting multiple times. This is a 1d restriction of the
shock-bubble interaction used in the original Ghost Fluid
paper~\cite{Fedkiw1999457}. The relativistic analogue used here has
initial data
\begin{equation}
  \label{eq:tests_flat_shock_slab_ID}
  \left\{ \begin{array}{clclclclc}
    \rhz = &  1.37795, \quad & v = & 0.17933, \quad & p = & 1.57, \quad
    & \gamma = & 1.4 \qquad & x < 0.25 \\
    \rhz = &  1, \quad  & v = & 0, \quad & p = & 1, \quad
    & \gamma = & 1.4 \qquad & 0.25 \le x < 0.45 \\
    \rhz = &  0.138, \quad  & v = & 0, \quad & p = & 1, \quad
    & \gamma = & 1.67 \qquad & 0.45 \le x < 0.55 \\
    \rhz = &  1, \quad  & v = & 0, \quad & p = & 1, \quad
    & \gamma = & 1.4 \qquad & x \ge 0.55.
  \end{array} \right. 
\end{equation}
The test is run to $t=0.8$.

\begin{figure}[htbp]
  \centering
  \includegraphics[width = 0.7\textwidth]{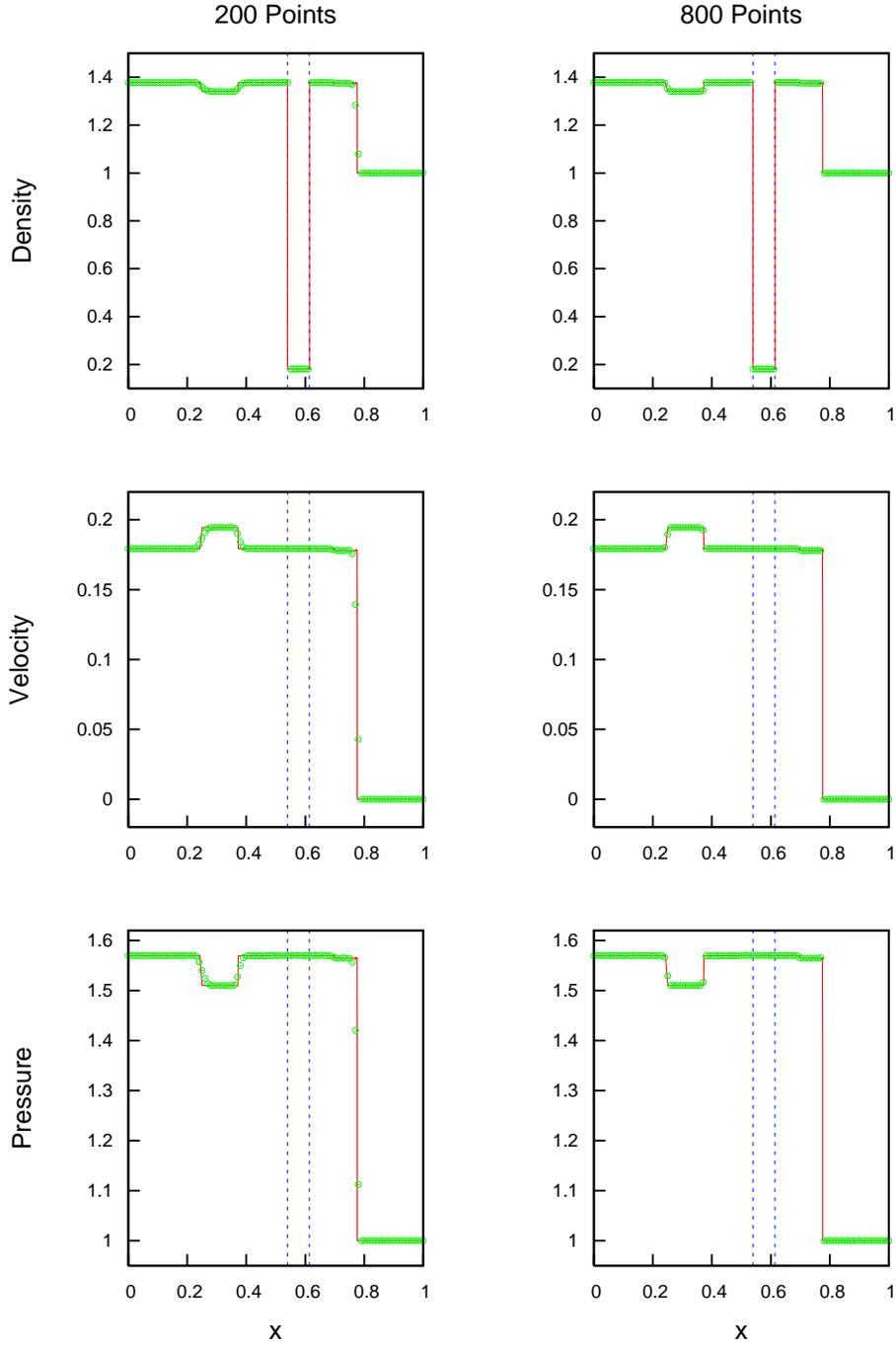}  
  \caption{The results for an isolated shock wave hitting a low
    density slab of material.  Multiple reflected and transmitted
    waves can be seen, as expected.  The movement and compression of
    the low density slab is also apparent. Velocity and pressure once
    again remain continuous at the interface as desired. TVD
    reconstruction using the MC limiter is used, and only 100 points
    are shown in each plot.}
  \label{fig:tests_flat_shock_slab}
\end{figure}
Results for this more complex interaction are shown in
figure~\ref{fig:tests_flat_shock_slab}. TVD reconstruction using the
MC limiter is used. The strong interacting features are captured, with
the interfaces moving independently in reaction to the shocks. No
oscillations occur, and the result converges with resolution.

\subsubsection{Perturbed shock}
\label{sec:tests_flat_ptb_shock}

A more complex example is that of a shock perturbed by a strong,
smooth feature. This test, suggested by~\cite{Dolezal1995266} and used
e.g.\ by~\cite{DelZanna02}, does not have an exact solution, so is
compared to a reference solution computed with 12800 points. We modify
this test to include an interface, using the initial data
\begin{equation}
  \label{eq:tests_flat_ptb_shock_ID}
  \left\{ \begin{array}{clclclclc}
    \rhz = &  5, \quad & v = & 0, \quad & p = & 50 ,
    \quad & \gamma = & 1.4 \qquad & x < 0.5 \\
    \rhz = &  2 + 0.3 \sin(50x), \quad  & v = & 0, \quad & p =
    & 5, \quad & \gamma = & 1.67 \qquad & x \ge 0.5
  \end{array} \right. .
\end{equation}
Again the domain is $x \in [0, 1]$ and the test is run to time $t =
0.35$.

\begin{figure}[htbp]
  \centering
  \includegraphics[width = 0.7\textwidth]{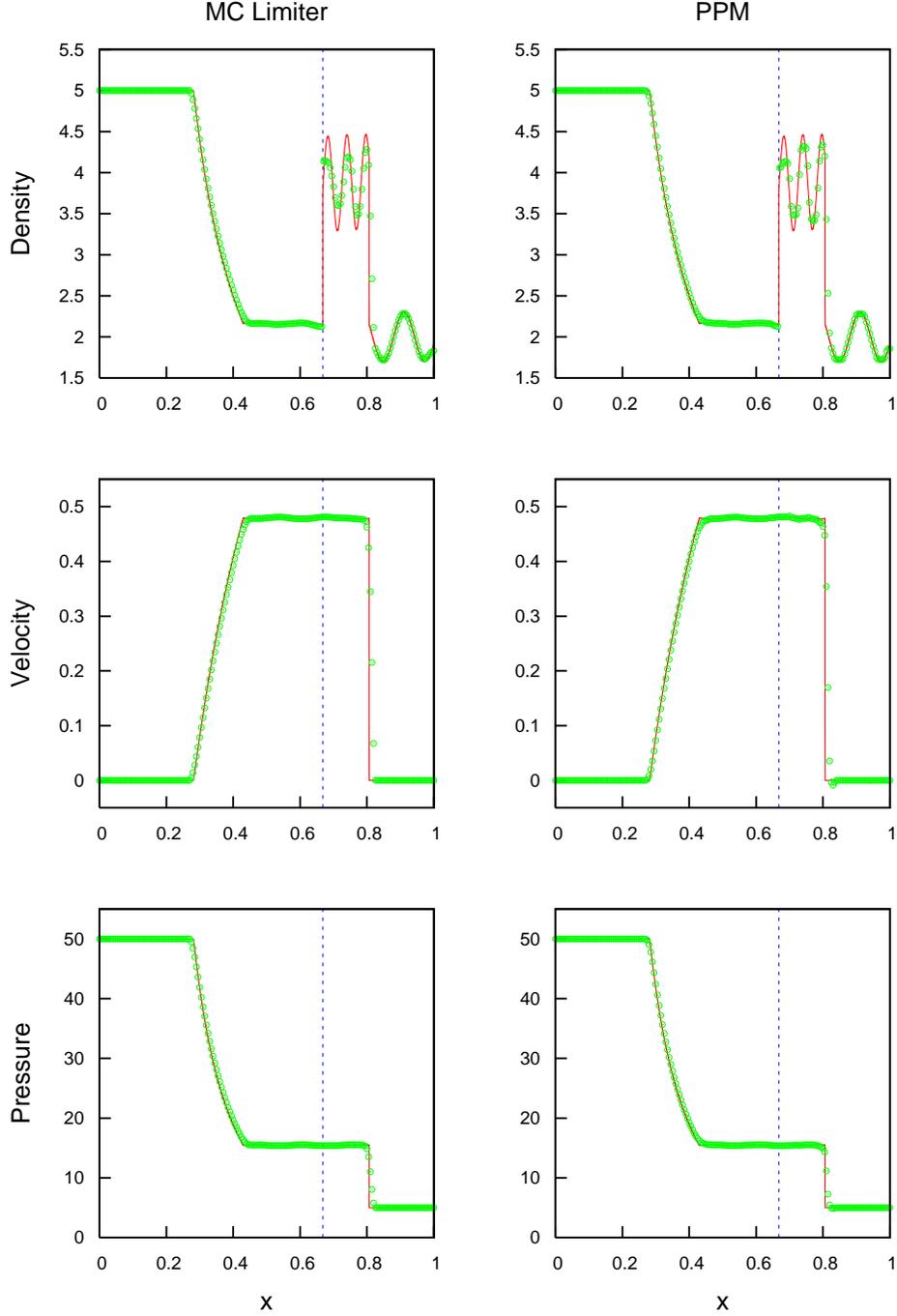}  
  \caption{Results for a test with a shock tube test with a sine wave
    density profile for 200 points.  The higher accuracy when using
    PPM is evident in the post shock density profile, with the sine
    wave being captured more accurately. The solid line shows a
    reference solution computed using 12800 points.}
  \label{fig:tests_flat_ptb_shock_200}
\end{figure}
\begin{figure}[htbp]
  \centering
  \includegraphics[width = 0.7\textwidth]{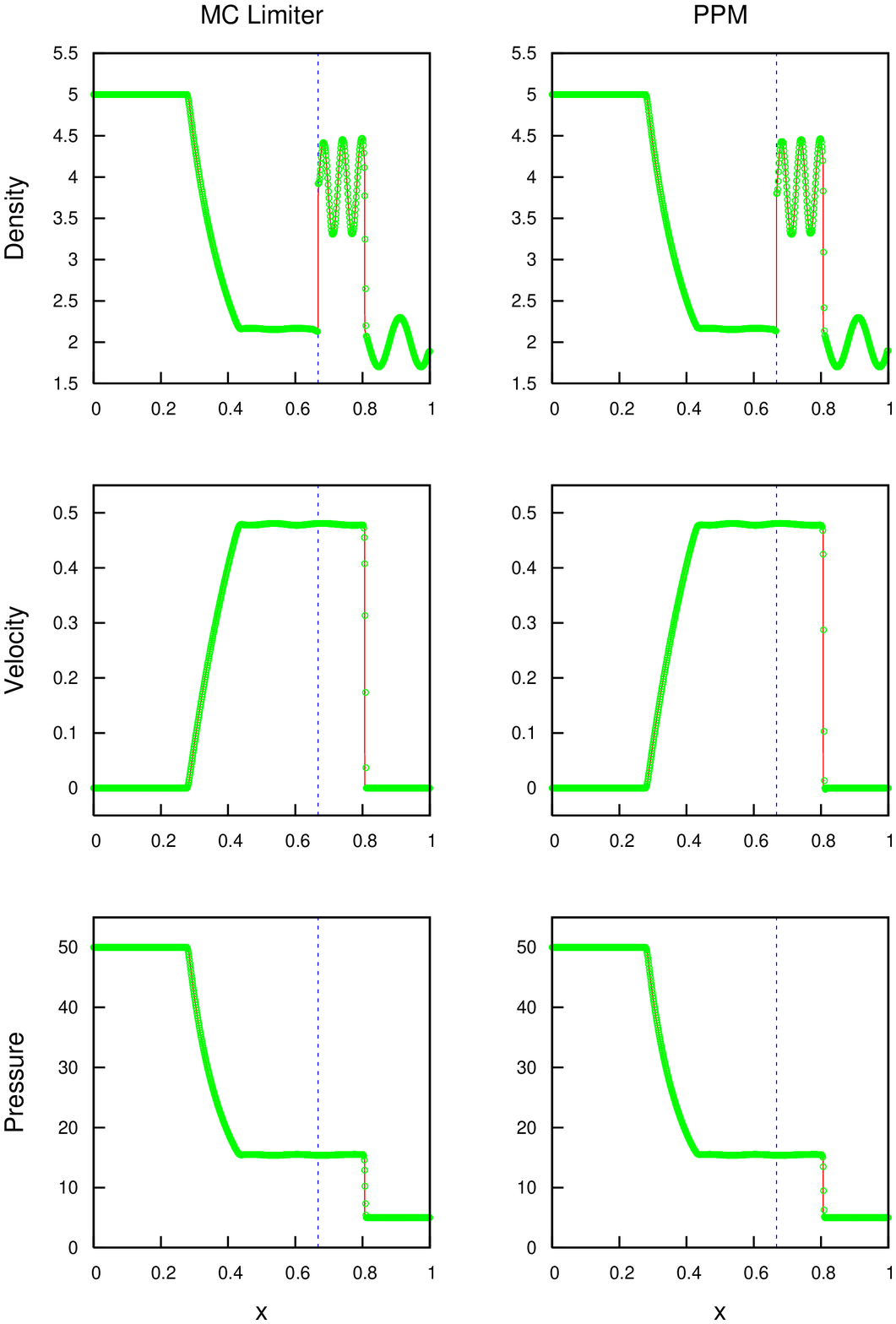}  
  \caption{Results for a test with a shock tube test with a sine wave
    density profile for 800 points.  In addition to greater accuracy
    for the maxima and minima of the sine wave, similar improvements
    in accuracy at the shock wave are visible for this test.} 
  \label{fig:tests_flat_ptb_shock_800}
\end{figure}
Results for this test are shown in
figures~\ref{fig:tests_flat_ptb_shock_200} and
\ref{fig:tests_flat_ptb_shock_800}. It is clear that the results are
converging to the correct solution irrespective of the numerical
method used. No spurious oscillations are introduced. However, it
appears that there is a slight discrepancy at the material
interface. In particular, the edges of the smooth feature do not
appear quite correct. We investigate this with another test.

\subsubsection{Moving sine wave}
\label{sec:tests_flat_moving_sine}

A final flat space test is adapted from a Newtonian test suggested
by~\cite{lombard:208}. A sine wave is advected in a surrounding
flow. The material interfaces should be stable, and so the evolution
should be trivial. The initial data used here is
\begin{equation}
  \label{eq:tests_flat_moving_sine_ID}
  \left\{ \begin{array}{clclclclc}
    \rhz = &  1, \quad & v = & 0.5, \quad & p = & 1 ,
    \quad & \gamma = & 1.4 \qquad & x < 0.16 \\
    \rhz = &  1 + 0.3 \sin(50(x-0.16)), \quad  & v = & 0.5, \quad & p =
    & 1, \quad & \gamma = & 1.67 \qquad & 0.16 \le x <
    0.537 \\ 
    \rhz = &  1, \quad  & v = & 0.5, \quad & p = & 1  ,
    \quad & \gamma = & 1.4 \qquad & x \ge 0.537
  \end{array} \right. .
\end{equation}
Again the domain is $x \in [0, 1]$ and the test is run to time $t =
0.4$.

\begin{figure}[htbp]
  \centering
  \includegraphics[width = 0.7\textwidth]{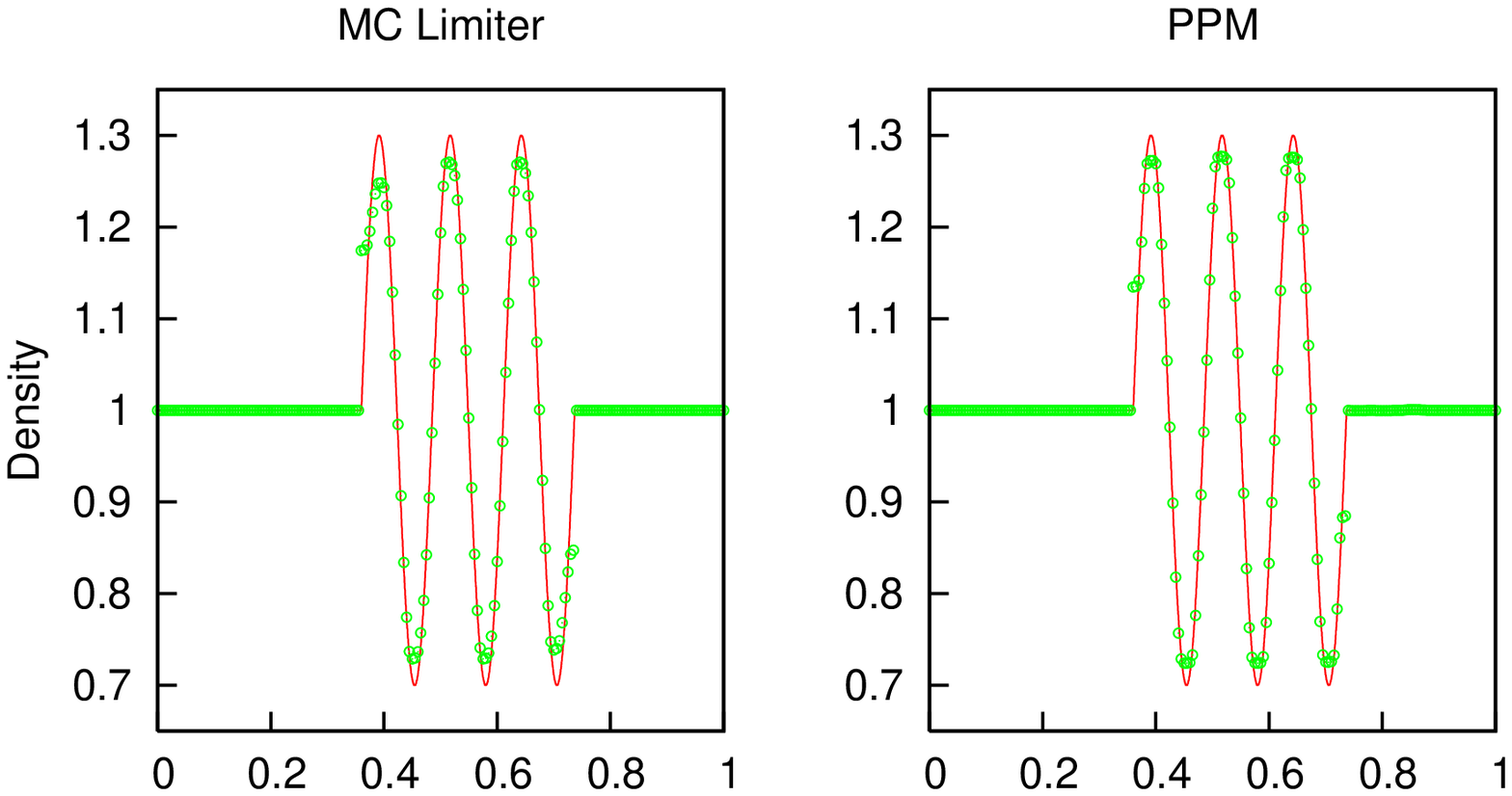}  
  \caption{An advected sine wave, evolved with a resolution of 200
    points.  Since the initial density profile is simply advected with
    the fluid velocity, an exact solution exists for this test.  The
    differences between the two reconstruction methods are evident,
    with the PPM test has more accurately capturing the sine wave.}
  \label{fig:tests_flat_moving_sine_200}
\end{figure}

\begin{figure}[htbp]
  \centering
  \includegraphics[width = 0.7\textwidth]{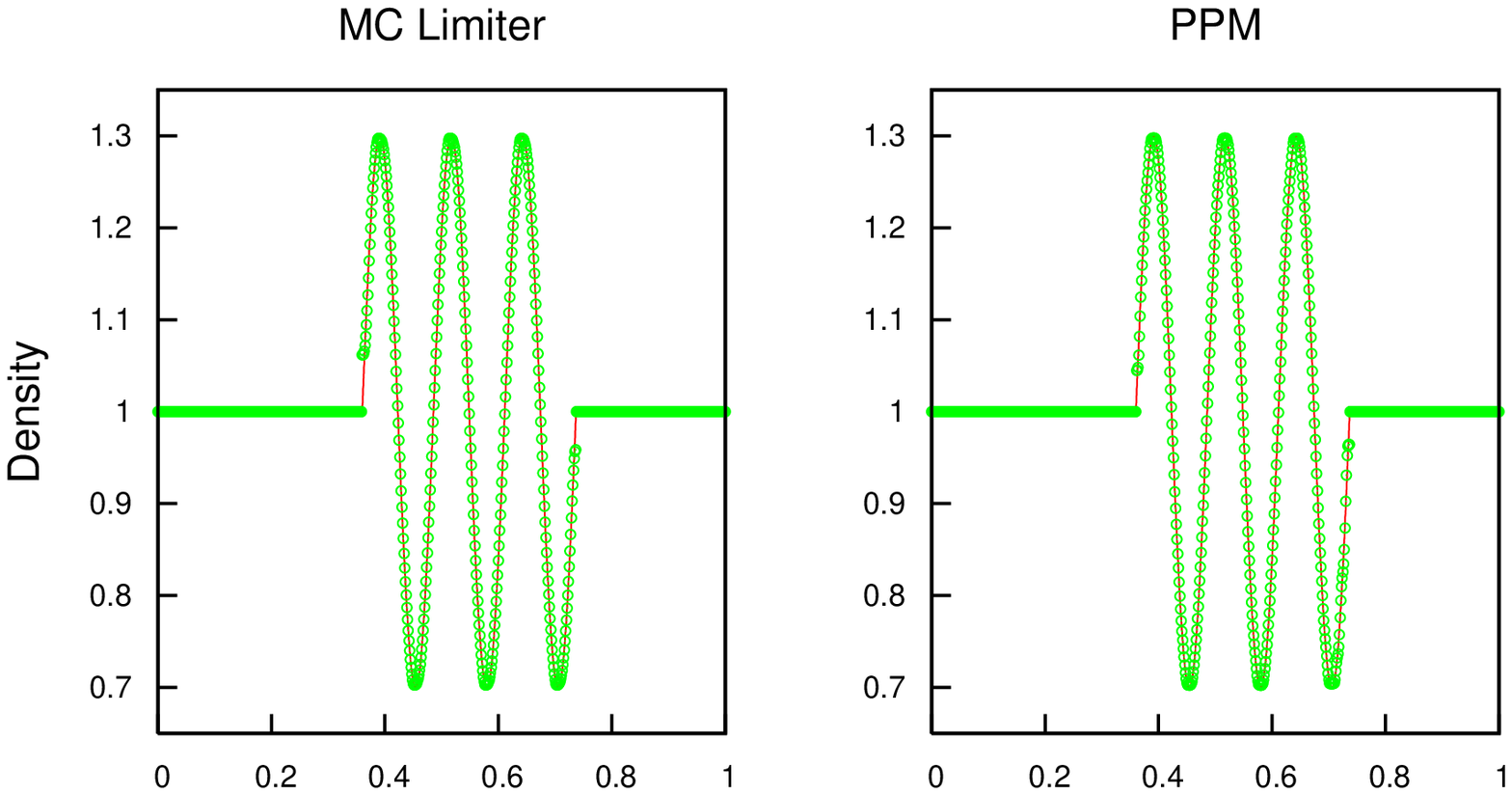}  
  \caption{An advected sine wave, evolved with a resolution of 800
    points.  The sine wave is now captured well with both techniques,
    though at the interface between the fluids there is still some
    error. Comparing to figure~\ref{fig:tests_flat_moving_sine_200}
    shows the (slow) convergence of this error.}
  \label{fig:tests_flat_moving_sine_800}
\end{figure}
Results for this test are shown in
figures~\ref{fig:tests_flat_moving_sine_200} and
\ref{fig:tests_flat_moving_sine_800}. The majority of the features are
well captured and no spurious oscillations are introduced at the
material interfaces. It is clear, however, that the accuracy suffers
near the interfaces due to the strong gradients there, and the overly
simple Ghost Fluid boundary condition. This converges with
resolution. We will see in the next section that in simple GR
simulations modelling neutron stars this is not a major concern.

\subsection{GR tests}
\label{sec:tests_gr}

In this section we look at tests in nonlinear general relativity,
focusing on tests relevant for neutron stars. We use the system of
equations outlined in section~\ref{sec:implement_form_hydro_sph}. All
of the tests look at static or nonlinearly perturbed TOV stars in
spherical symmetry. The initial data is always constructed with a
polytropic EOS
\begin{equation}
  \label{eq:test_eos_polytrope}
  p = K \rho_0^{\gamma}
\end{equation}
where $K$ is the polytropic constant, and evolved with the ideal fluid
EOS of equation~(\ref{eq:EOS}), where $\gamma$ may vary in space.

\subsubsection{Static stars}
\label{sec:static}

As a reference solution we construct and evolve a standard static
single component star. The initial data is determined by the initial
central density $\rho_0(t=0,r=0)$ denoted $\rho_c$, and the EOS. Here
we use the values
\begin{equation}
  \label{eq:tests_1_comp_ID}
  \rho_c = 1.28 \times 10^{-3}, \qquad \gamma = 2, \qquad K = 100.
\end{equation}
We then evolve this using the TVD method as described in
section~\ref{sec:implement_numerics}. 
\begin{figure}
  \centering
  \includegraphics[angle=270,width = 0.85\textwidth]{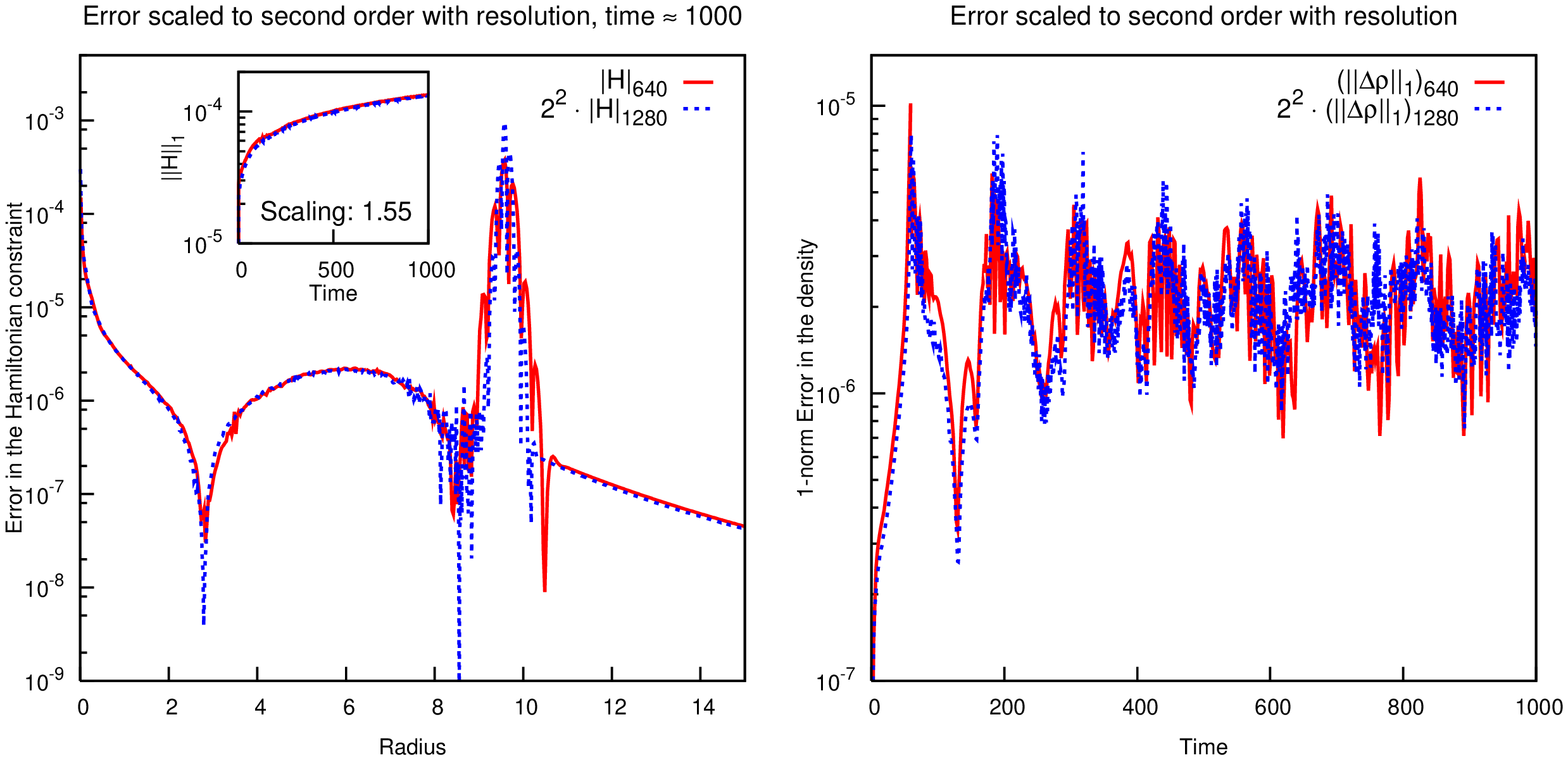}
  \caption{A single component TOV star is evolved as a reference
    case. The error in the Hamiltonian constraint is shown as a
    function of space at late time in the left panel, with the
    time-dependent behaviour of the 1-norm shown in the
    inset. Simulations with 640 and 1280 points are shown, with the
    behaviour in space scaled for second order convergence. The
    behaviour of the norm is scaled for convergence order 1.55. Second
    order convergence is observed nearly everywhere. The surface,
    located near $r \sim 10$, is the clear location where the order of
    convergence is reduced, as expected. This leads to the reduced
    order of convergence in the 1-norm. In the right panel the 1-norm
    of the density errors is shown, scaled for second order
    convergence. Here the impact of the surface errors is not so
    pronounced, so second order convergence in the norm is
    approached.}
  \label{fig:GR_1comp}
\end{figure}
The errors in the evolution are shown in
figure~\ref{fig:GR_1comp}. Second order convergence is seen over the
majority of space at late time. However, near the surface of the star
convergence is lost, both due to the ``clipping'' inherent in HRSC
methods, and due to the atmosphere algorithm used here. This is the
dominant factor in reducing the order of convergence of the norms.

As a first test of the accuracy of the Ghost Fluid algorithm in GR we
then introduce a ``trivial'' interface into the TOV star at coordinate
location $r = 3.015$. On either side of this interface the EOS is the
same as that given in equation~(\ref{eq:tests_1_comp_ID}). We then
evolve, imposing the Ghost Fluid boundary condition at the
interface. It should be noted that in this simulation, although the
level set is evolved the interface inferred from the zero of the level
set remains within the same grid cell at all times.
\begin{figure}
  \centering
    \includegraphics[angle=270,width = 0.85\textwidth]{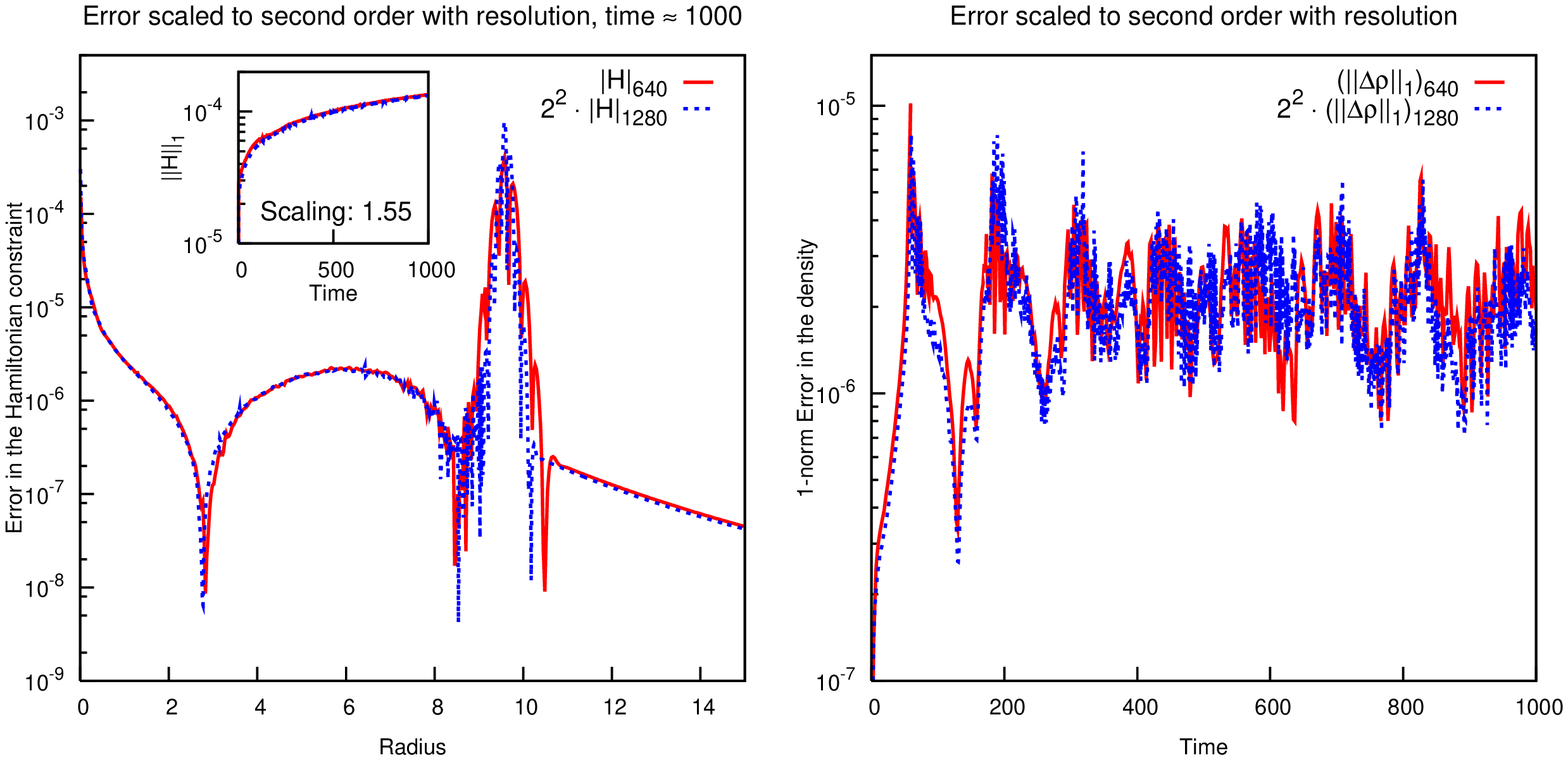}
    \caption{A single component TOV star with an interface
      artificially introduced at $r = 3.015$ is evolved. These results
      should be compared with the reference solution in
      figure~\ref{fig:GR_1comp}. The accuracy and convergence
      properties are nearly identical to the reference solution.}
  \label{fig:GR_2comp_equal}
\end{figure}
As can be seen from figure~\ref{fig:GR_2comp_equal}, by comparing to
the reference solution of figure~\ref{fig:GR_1comp}, the Ghost Fluid
method makes virtually no difference to the accuracy of the
results. The convergence properties are nearly identical, and the
effect of the interface is very difficult to detect in the errors.

\begin{figure}
  \centering
  \includegraphics[angle=270,width = 0.85\textwidth]{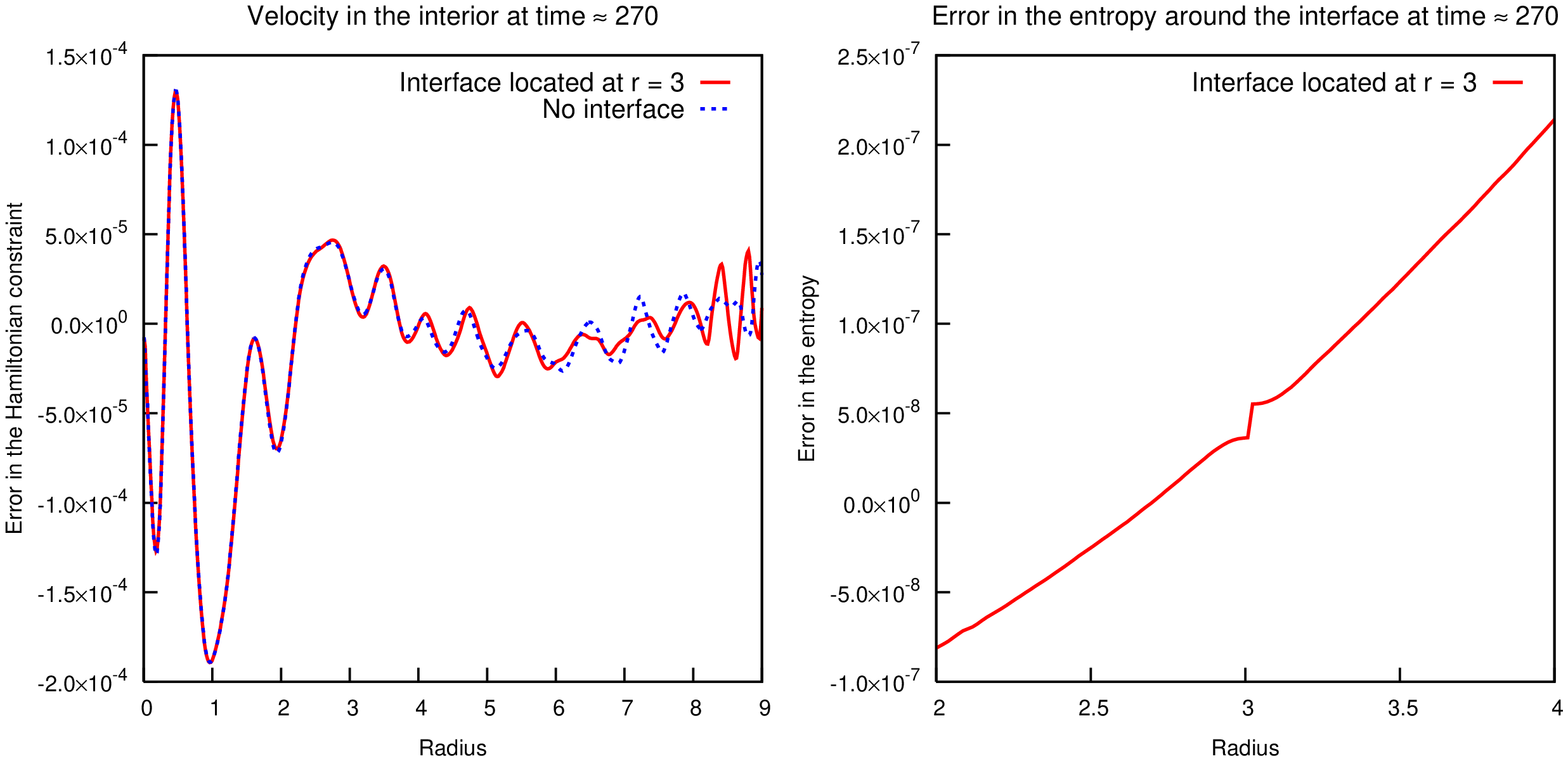}
  \caption{A closer look at the evolution of the single component TOV
    star with the artificial interface at $r \simeq 3$. Only the
    simulation using 1280 points is shown. In the left panel the
    radial 3-velocity component is shown. As expected, the velocity is
    continuous across the interface. The difference between this
    evolution and the reference solution are due to interactions
    between the errors introduced by the surface and the interface,
    and are strongest near the surface. In the right panel the change
    in the entropy between the initial data and $t \simeq 270$ is
    shown. As the entropy is extrapolated across the interface this
    shows the jump induced by the Ghost Fluid boundary condition;
    however, the size of this error is negligible.}
  \label{fig:GR_2comp_equal_v_s}
\end{figure}
To look at the effect of the Ghost Fluid boundary condition in more
detail, in figure~\ref{fig:GR_2comp_equal_v_s} we show the difference
in the radial 3-velocity component between the reference evolution
without the interface and the evolution containing the interface. The
effect of the interface is not directly obvious in the plot. The
velocity should be continuous at the interface, and clearly is. The
differences between the two evolutions appear near the surface of the
star, which is where the velocity is largest. This indicates that the
differences are due to minor reflections from the interface which are
clearly at a very low level. In addition we look at the change in the
entropy during the evolution. The entropy is extrapolated by the Ghost
Fluid algorithm, so should show the effect of the boundary
condition. The initial entropy is ${\cal O}(1)$, so whilst the effect
of the boundary condition is clear in inducing a small jump at the
interface, the magnitude of the effect at $\sim 10^{-7}$ is clearly
negligible.

\begin{figure}
  \centering
    \includegraphics[angle=270,width = 0.85\textwidth]{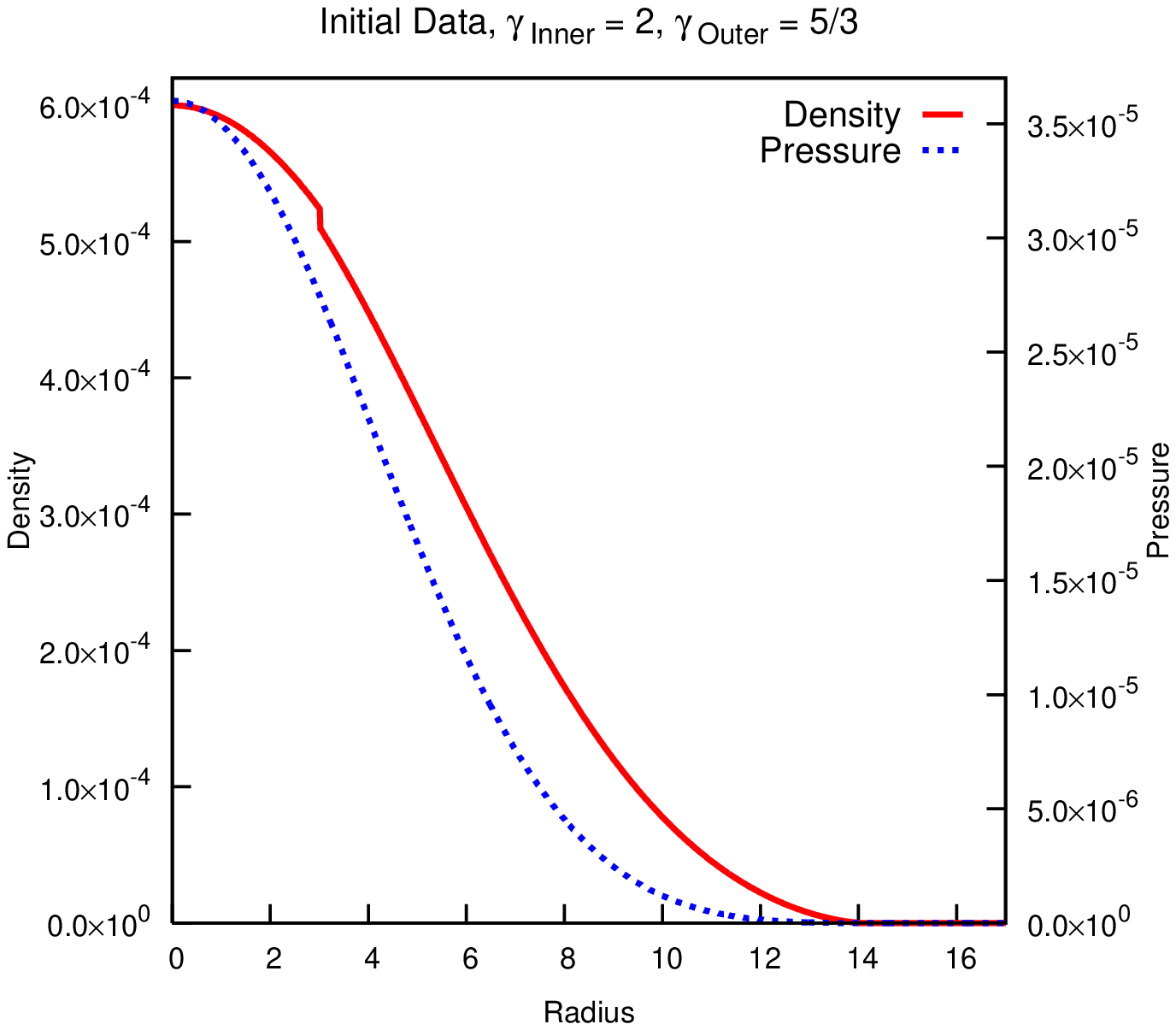}  
  \caption{The initial data used for the evolution of a two component
    star. The outer fluid uses a softer EOS than the reference
    solution, leading to a larger radius. The interface at $r \simeq
    3$ is clearly seen from the density jump.}
  \label{fig:GR_2comp_diff_ID}
\end{figure}

We then move on to look at static stars containing a genuine interface
between two different components. We have studied a number of
different configurations, with the qualitative behaviour being similar
in all cases. As an illustration we use the initial data
\begin{equation}
  \label{eq:tests_2_diff_comp_ID}
  \left\{ \begin{array}{clclclc}
    \rho_c = &  6 \times 10^{-4}, \quad & \gamma = & 2, \quad & K =
    & 100 , \qquad & r < 3.015 \\
    &  \quad  & \gamma = & \frac{5}{3}, \quad & K =
    & 11.17, \qquad & r \ge 3.015 
  \end{array} \right. .
\end{equation}
This ensures the TOV is stable, and that pressure and its first
derivative are continuous at the interface. This initial data is
illustrated in figure~\ref{fig:GR_2comp_diff_ID}.
\begin{figure}
  \centering
    \includegraphics[angle=270,width = 0.85\textwidth]{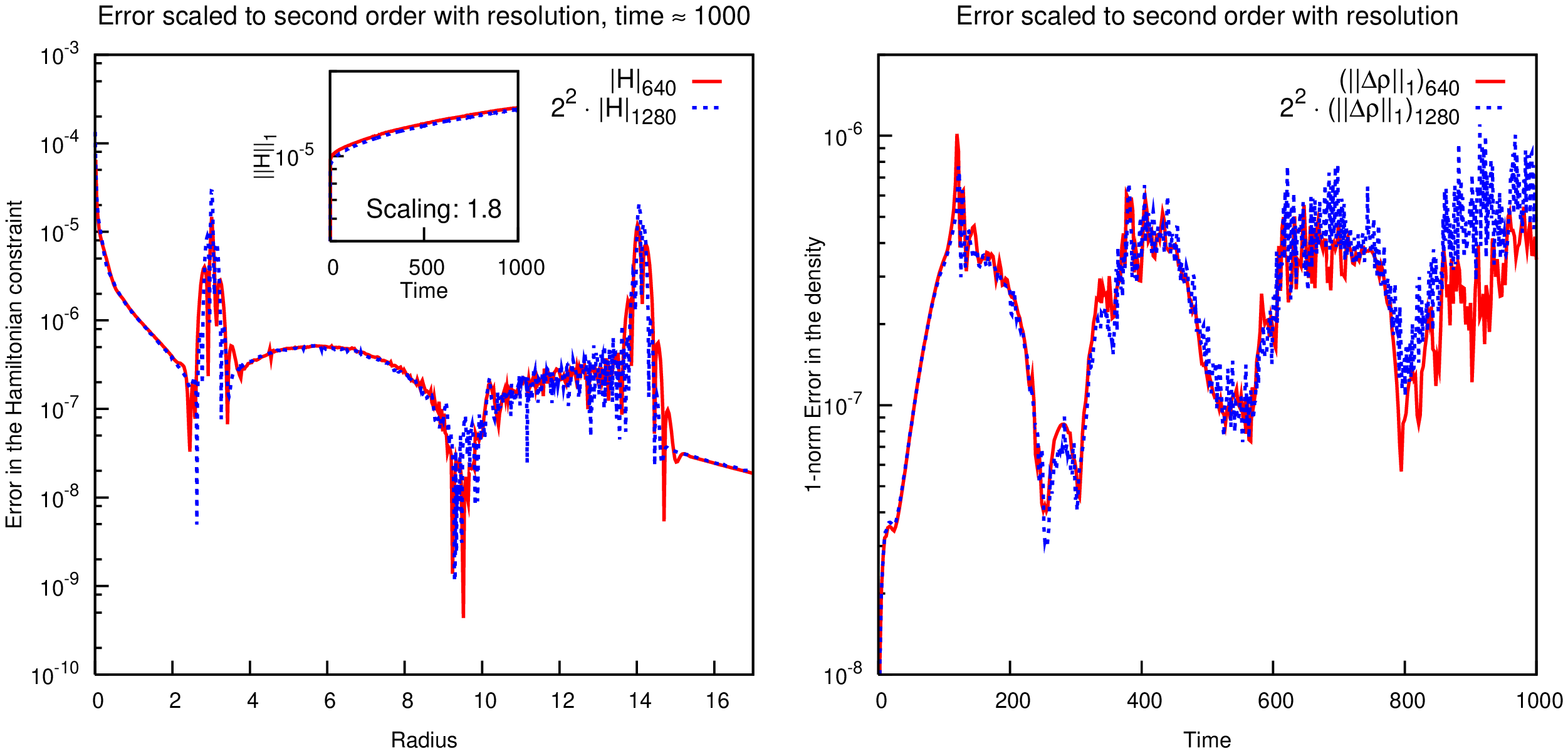}  
    \caption{Errors introduced in evolving a two component star. This
      should be compared to the reference solution in
      figure~\ref{fig:GR_1comp} and the artificial interface solution
      in figure~\ref{fig:GR_2comp_equal}. The interface at $r \simeq
      3$ now has a visible effect in the errors, but these are still
      manageable and converge. Note that at these resolutions the
      convergence of the norm of the constraints is actually better
      than in the reference solution, as the errors in the interior
      (which converge at second order) are larger due to the
      non-trivial interface, whilst those near the surface (which
      converge at less than second order) are smaller due to the
      softer EOS. The errors in the density show good convergence
      until late times when the errors near the surface become
      sufficiently large.}
  \label{fig:GR_2comp_diff}
\end{figure}
The result of the evolution is shown in
figure~\ref{fig:GR_2comp_diff}. In this case we do see an effect in
the Hamiltonian constraint from the interface at $r = 3.015$. This
leads to larger errors within the star (at fixed resolution). However,
these errors appear to converge at second order (except for the points
directly next to the interface). This means that, for these
resolutions, the convergence of the constraint error is actually
better than for the reference solution shown in
figure~\ref{fig:GR_1comp}. This is because the error at the surface,
which does not converge at second order due to the atmosphere
algorithm and the inherent properties of HRSC methods, contributes
less (relatively) to the total error. In contrast, the errors in the
density are (relatively) larger at the surface as we are using a lower
central density and a softer EOS for the outer fluid. Hence at late
times as the surface errors increase the effects are seen in the
errors of the density, which drifts away from perfect second order
convergence.

\subsubsection{Perturbed star}
\label{sec:perturbed}

The tests in section~\ref{sec:static} show that there is no problem in
extending the level set method augmented with the Ghost Fluid boundary
condition to GR when studying simple solutions. However, we really
require that these methods will work in strongly nonlinear situations
which will include interface motion and shock formation and
interaction. In order to look at this in a 1+1 context we take a two
component TOV star and give a strong nonlinear perturbation designed
both to move the interface and to produce shocks propagating within
the fluid. Due to the limitations of our atmosphere algorithm the
results will not be trustworthy when the shocks start to reach the
surface of the star, so we keep the evolutions relatively
short. However, these evolutions are sufficient to validate the
approach that we use here.

The initial data used is based on a two component TOV star with
parameters
\begin{equation}
  \label{eq:tests_2_comp_pert_ID}
  \left\{ \begin{array}{clclclc}
    \rho_c = &  1.28 \times 10^{-3}, \quad & \gamma = & 2, \quad & K =
    & 100 , \qquad & r < 3.015 \\
    &  \quad  & \gamma = & 1.9, \quad & K =
    & 51.57, \qquad & r \ge 3.015 
  \end{array} \right. .
\end{equation}
The interior is then perturbed using
\begin{equation}
  \label{eq:tests_2_comp_pert}
  \rho_0 = \left(\rho_0\right)_{\text{TOV}} \left( 1 + h(r) \right), \quad
  p = p_{\text{TOV}} \left( 1 + h(r) \right),
\end{equation}
where
\begin{equation}
  \label{eq:tests_2_comp_pert_h}
  h(r) =
  \begin{cases}
    \frac{1}{20}
    \left( 1 - \tanh \left[ 50 (r - 2) \right] \right) & r < 2.5, \\
    0 & r \ge 2.5.
  \end{cases}
\end{equation}
\begin{figure}
  \centering
  \includegraphics[angle=270,width = 0.85\textwidth]{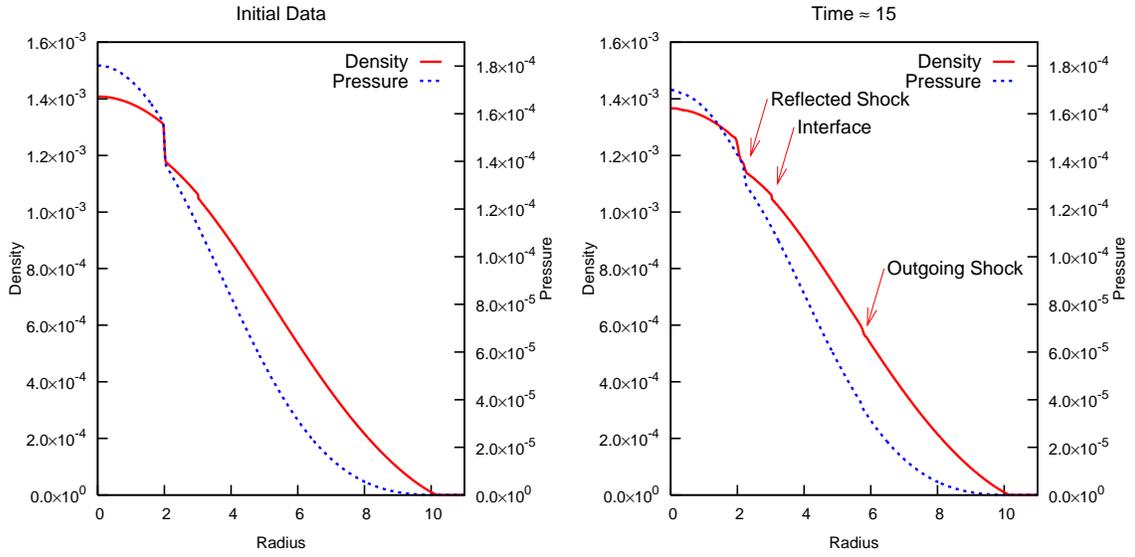}
  \caption{The left panel shows the initial data for the nonlinearly
    perturbed two component star. The interior perturbation near $r
    \simeq 2$ is not quite a shock but will trigger an outgoing wave
    that steepens into a shock before passing through the interface at
    $r \simeq 3$. It will also produce a strong ingoing wave that will
    reflect from the origin, producing another shock. The right panel
    shows the evolved at a time when the initial outgoing shock has
    passed through the interface but before the reflected shock has
    had time to reach it.}
  \label{fig:GR_2comp_shock_ID}
\end{figure}
This is illustrated in the left panel of
figure~\ref{fig:GR_2comp_shock_ID}. The strong perturbation leads to a
very steep feature near $r \simeq 2$ that will form a shock and move
the interface. As seen in the right panel of
figure~\ref{fig:GR_2comp_shock_ID}, there is an additional shock
formed by the ``ingoing'' piece of the initial perturbation reflecting
from the origin; at the time of this plot it has not reached the
interface. 

\begin{figure}
  \centering
  \includegraphics[angle=270,width = 0.85\textwidth]{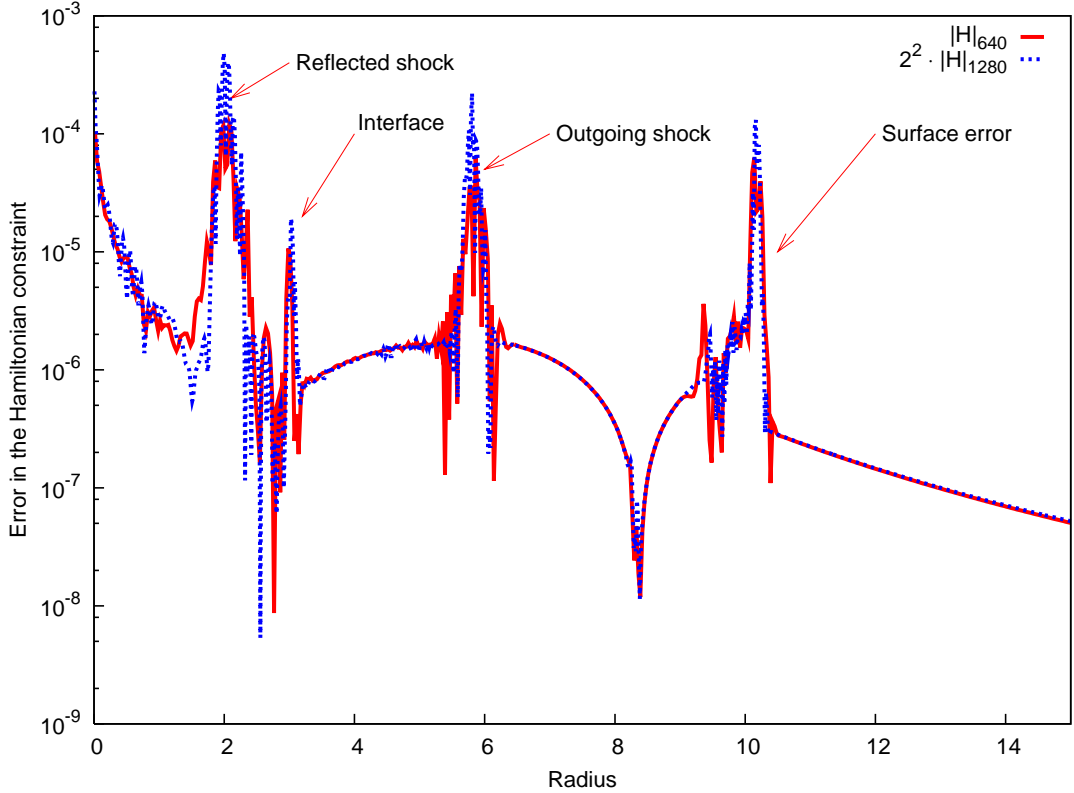}
  \caption{Constraint errors in the evolution of the nonlinearly
    perturbed star. The curves are scaled for second order
    convergence, which is seen away from the surface and the shocks.}
  \label{fig:GR_2comp_shock}
\end{figure}
The constraint errors in the evolution of the nonlinearly perturbed
star are shown in figure~\ref{fig:GR_2comp_shock}. As expected we now
see a large number of features indicated on the plot. As in the
reference solution in figure~\ref{fig:GR_1comp} we see the error from
the surface, which is sizeable and does not converge cleanly at second
order. As in the static two component test in
figure~\ref{fig:GR_2comp_diff_ID} we see an error at the interface,
which in this case is relatively small and converges except at a few
points directly neighbouring the interface. Finally we see strong
features at the two shocks where, due to the nature of HRSC methods,
the order of accuracy is degraded to first order.

\begin{figure}
  \centering
  \includegraphics[angle=270,width = 0.85\textwidth]{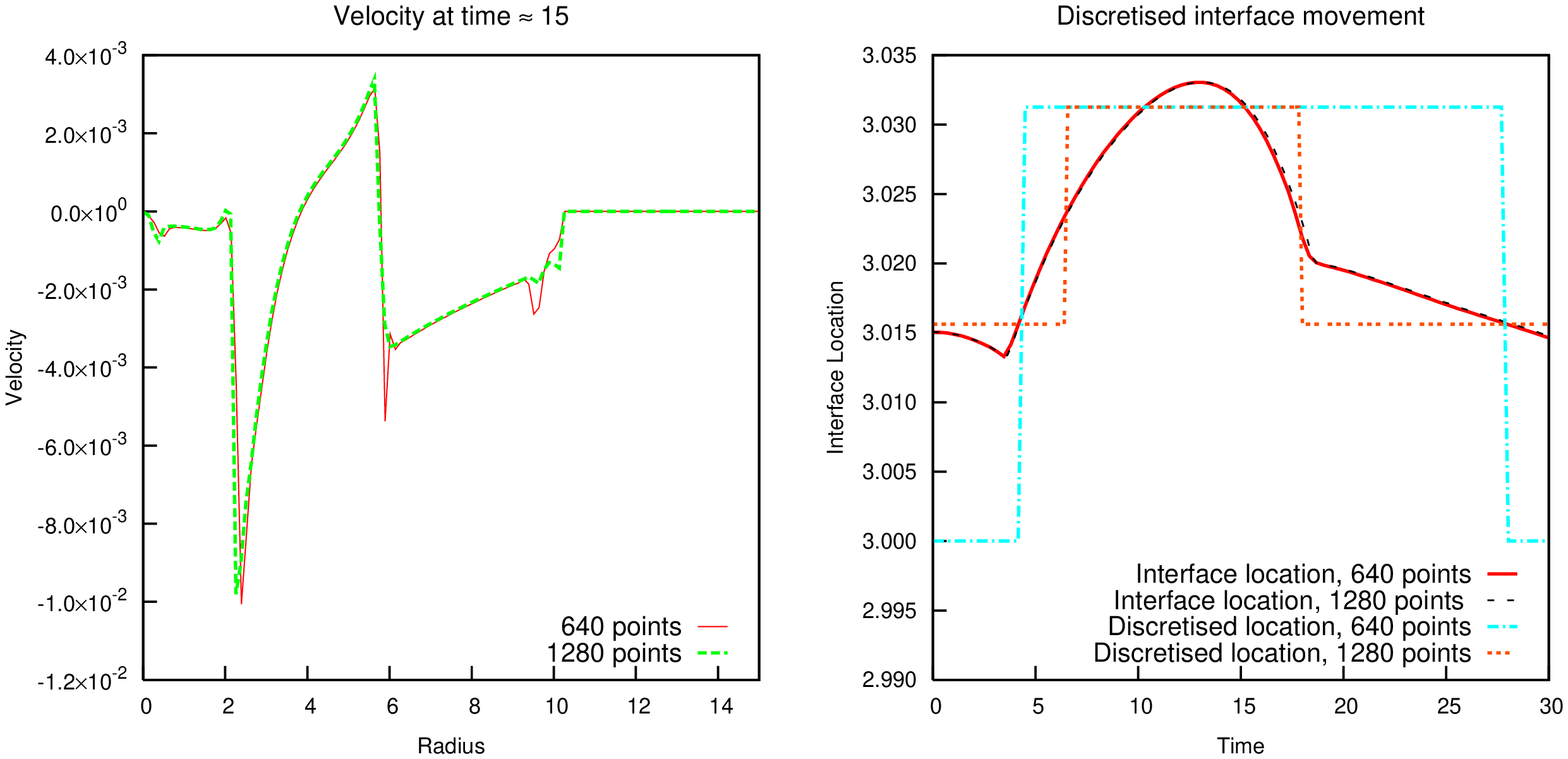}
  \caption{The left panel shows the radial component of the 3-velocity
    for the nonlinearly perturbed star, indicating the locations of
    the shocks. By this point the interface, initially at $r = 3.015$,
    has moved outwards and has started to move back in again. As
    expected the velocity is smooth at the interface and no spurious
    oscillations have been generated by the interaction with the
    outgoing shock. The right panel indicates the interface location
    as a function of time, showing its coordinate location as inferred
    from the level set, and the computational coordinate of the centre
    of the cell within which the interface is found. We see that even
    with this strong perturbation the motion is small but similar for
    all resolutions.}
  \label{fig:GR_2comp_shock_interface}
\end{figure}
Figure~\ref{fig:GR_2comp_shock_interface} shows the motion of the
interface and also shows that no spurious oscillations occur due to
the interaction between the outgoing shock and the interface. The left
panel shows the velocity at a time when the outgoing shock has passed
through the interface located at $r \simeq 3$. The velocity is
continuous with no spurious oscillations, despite the interaction with
the shock. The right panel shows the motion of the
interface. Initially it falls in, due to the additional matter in the
interior. It then interacts with the shock, being pushed outwards and
jumping between cells, before starting to fall back again, with the
additional interaction with the reflected shock at $t \simeq 18$ being
insufficient to stop its inward motion.


\section{Conclusions} 
\label{sec:concl}


In this paper we have considered models of relativistic matter
containing sharp interfaces and their particular importance and
implementation for numerical simulations. We expect that simulations
of realistic models of neutron stars, as required for gravitational
wave template calculations, will contain features that will be too
sharp to be resolved as a smooth feature using reasonable numerical
resolution. It follows that models incorporating sharp interfaces and
techniques for evolving them cleanly will be required.

The use of level set techniques to capture and evolve the interfaces
that might form and disappear through mergers is already well known in
numerical relativity, particularly in the context of event horizon
finders. Their use here is natural and simple, and should allow the
techniques investigated here to easily be extended to more spatial
dimensions. 

However, the description of the interface location is not sufficient.
We must impose a boundary condition at the interface, which will
complete the description of the continuum model.  The Ghost Fluid
boundary condition used here has the advantage of simplicity, which
makes it ideal for a proof-of-principle test such as this. However, it
is questionable if this condition is sufficient for all situations,
and more advanced techniques may be required with more complex EOSs
than those considered here.

The results of section~\ref{sec:tests} illustrate the advantages of
the combined level set and Ghost Fluid method. It is simple, and for
strong shocks or smooth flow interacting with interfaces it gives good
results. We have shown that it directly extends to GR, and that for
mildly perturbed neutron star tests the results are stable,
convergent, and as accurate as the numerical method employed. However,
when there is a long-lasting, strong feature that interacts with the
interface, the test shown in section~\ref{sec:tests_flat_moving_sine}
illustrates the limitations of the technique. This may indicate the
need for better boundary conditions at the interface before high
accuracy simulations for gravitational wave extraction can be
attempted for complex neutron star models.


\[ \]

{\bf Acknowledgements:}

IH was supported by the STFC rolling grant PP/E001025/1 and Nuffield
Foundation grant NAL/32622. SM was supported by an EPSRC studentship.


\appendix

\section{Colour functions}
\label{sec:appendix_colourfn}

At first glance it would appear that the simple tests shown in this
paper are solved with an excessively complex numerical method. From
the simplest point of view, our system of equations describing
relativistic hydrodynamics has simply been augmented with an equation
of motion for the EOS parameter $\gamma$. This can be described by a
``colour function'' and advected using standard HRSC methods if
written in the form
\begin{equation}
  \label{eq:colour_fn1}
  \partial_t \left( \rho_0 \gamma \right) + \partial_i \left( \rho_0
    \gamma v^i \right) = 0.
\end{equation}
In fact any parameter $\kappa$ required by the EOS can be advected
using an equation of this form.

However, it is well known in the CFD literature that this na{\"\i}ve
approach fails for the simplest interfaces. 
\begin{figure}
  \centering
  \includegraphics[angle=270,width=0.85\textwidth]{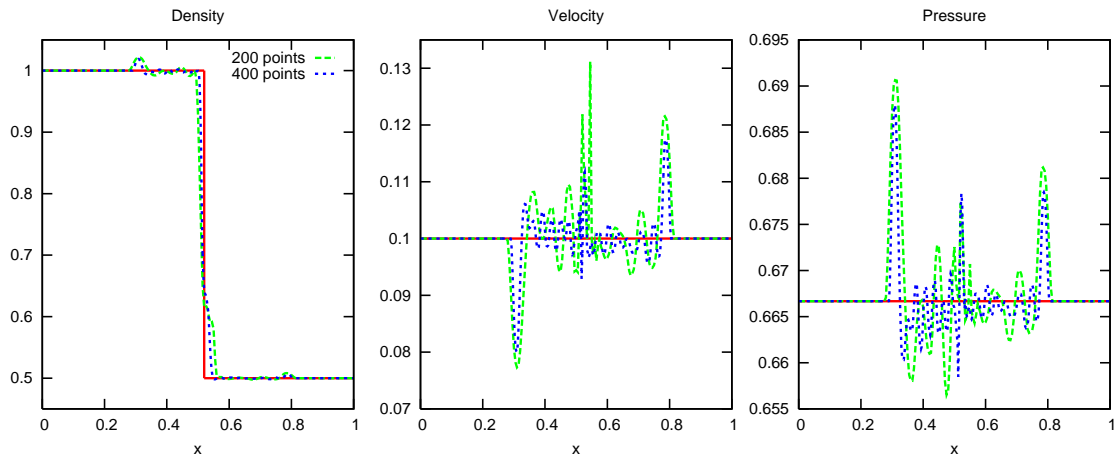}
  \caption{A stable contact discontinuity, initially at $x = 0.5$,
    with the fluid to the left having $\gamma = 5/3$ and the fluid to
    the right having $\gamma=4/3$ is advected to the right with speed
    $0.1$. The initial data is set such that the interface should
    propagate unchanged. The plot shows this initial data evolved
    using two resolutions and standard HRSC methods to time $t=0.2$
    using the colour function approach, with the exact solution given
    by the solid line. Spurious oscillations are introduced at the
    interface. These oscillations do not converge with resolution,
    indicating a failure of the approach.}
  \label{fig:colour_fn1}
\end{figure}
Figure~\ref{fig:colour_fn1} shows this approach failing for a stable
material interface. The interface is set at $x = 0.5$, with initial data
\begin{equation}
  \label{eq:colour_fn_ID}
  \left( \rho_0, v, p, \gamma \right) =
  \begin{cases}
    \left( 1, 0.1, \tfrac{2}{3}, \tfrac{5}{3} \right) & x < 0.5, \\
    \left( \tfrac{1}{2}, 0.1, \tfrac{2}{3}, \tfrac{4}{3} \right) & x
    \ge 0.5,
  \end{cases}
\end{equation}
leading to a stable interface propagating to the right. The standard
conservation law form, as in
section~\ref{sec:implement_form_hydro_flat}, augmented with the colour
function equation~(\ref{eq:colour_fn1}), is evolved using the simple
TVD HRSC method used in the rest of the paper. This is considerably
simpler than the level set method, but as can be seen from the figure
fails utterly to deal with this simple interface. Spurious
oscillations are introduced. These oscillations are sizeable and do
not converge with resolution. Clearly there is a failure of the
approach, which is due to the formulation of the problem.
\begin{figure}
  \centering
  \includegraphics[angle=270,width=0.85\textwidth]{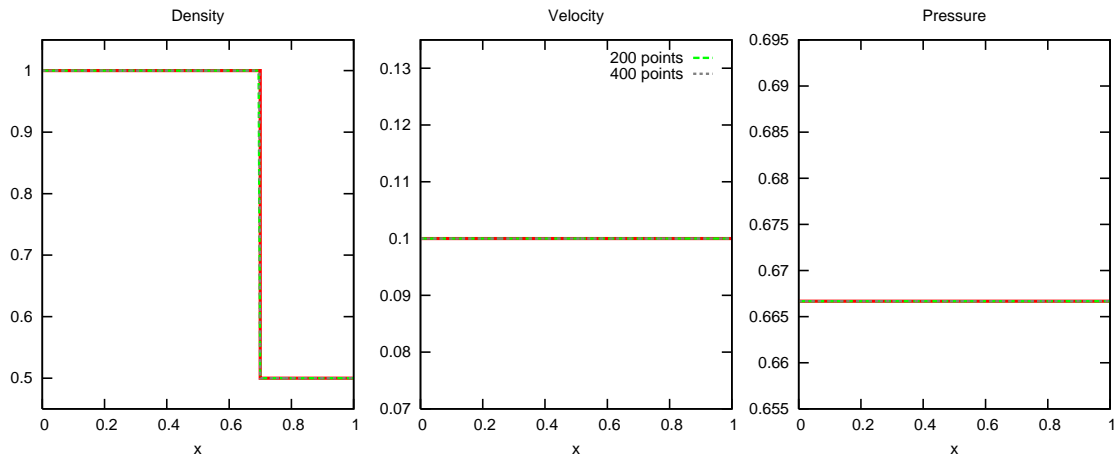}
  \caption{The stable contact discontinuity test evolved to time $t =
    2$ using the level set method with the Ghost Fluid boundary
    condition. Comparing to figure~\ref{fig:colour_fn1} we see that
    there are no spurious oscillations and the interface is captured
    perfectly at all resolutions.}
  \label{fig:colour_fn2}
\end{figure}
The level set method combined with the Ghost Fluid boundary condition
deals with this situation with no problem whatsoever, as illustrated
in figure~\ref{fig:colour_fn2}.

The reasons for the failure of the colour function approach have been
studied in detail by a number of authors. As explained
by~\cite{Abgrall2001594} the key is that the inherent ``smearing'' of
the HRSC scheme will effectively ``mix'' the fluids either side of the
interface in a narrow region. The detailed way that the mixing occurs
will depend on the resolution and the numerical method, but it will
always occur. However, without imposing further conditions it is clear
that this mixing will not be \emph{thermodynamically consistent}. That
is, the \emph{numerical} mixing will lead to spurious generation of
entropy at the interface, creating the oscillations observed. Whilst
it is possible to construct numerical methods that are
thermodynamically consistent (e.g.~\cite{Wang2004528}; for example,
for the EOS used here evolving the EOS parameter $\kappa = 1 / (\gamma
- 1)$ is consistent, as shown by~\cite{Abgrall2001594}) this may not
always be straightforward, and the problem of the length over which
the interface is smeared, as described in the introduction, remains.



\bibliographystyle{apsrev} 
\bibliography{GF}

\begin{thebibliography}{29}
\expandafter\ifx\csname natexlab\endcsname\relax\def\natexlab#1{#1}\fi
\expandafter\ifx\csname bibnamefont\endcsname\relax
  \def\bibnamefont#1{#1}\fi
\expandafter\ifx\csname bibfnamefont\endcsname\relax
  \def\bibfnamefont#1{#1}\fi
\expandafter\ifx\csname citenamefont\endcsname\relax
  \def\citenamefont#1{#1}\fi
\expandafter\ifx\csname url\endcsname\relax
  \def\url#1{\texttt{#1}}\fi
\expandafter\ifx\csname urlprefix\endcsname\relax\def\urlprefix{URL }\fi
\providecommand{\bibinfo}[2]{#2}
\providecommand{\eprint}[2][]{\url{#2}}

\bibitem[{\citenamefont{Font}(2008)}]{lrr-2008-7}
\bibinfo{author}{\bibfnamefont{J.~A.} \bibnamefont{Font}},
  \bibinfo{journal}{Living Reviews in Relativity} \textbf{\bibinfo{volume}{11}}
  (\bibinfo{year}{2008}),
  \urlprefix\url{http://www.livingreviews.org/lrr-2008-7}.

\bibitem[{\citenamefont{Abbott et~al.}(2009)}]{LIGOReview}
\bibinfo{author}{\bibfnamefont{B.~P.} \bibnamefont{Abbott}}
  \bibnamefont{et~al.}, \bibinfo{journal}{Reports on Progress in Physics}
  \textbf{\bibinfo{volume}{72}}, \bibinfo{pages}{076901 (25pp)}
  (\bibinfo{year}{2009}),
  \urlprefix\url{http://stacks.iop.org/0034-4885/72/076901}.

\bibitem[{\citenamefont{Acernese et~al.}(2006)}]{Acernese2006}
\bibinfo{author}{\bibfnamefont{F.}~\bibnamefont{Acernese}}
  \bibnamefont{et~al.}, \bibinfo{journal}{Class. Quantum Grav.}
  \textbf{\bibinfo{volume}{23}}, \bibinfo{pages}{S635} (\bibinfo{year}{2006}).

\bibitem[{\citenamefont{Andersson et~al.}(2005)\citenamefont{Andersson, Comer,
  and Glampedakis}}]{Andersson:2004aa}
\bibinfo{author}{\bibfnamefont{N.}~\bibnamefont{Andersson}},
  \bibinfo{author}{\bibfnamefont{G.~L.} \bibnamefont{Comer}}, \bibnamefont{and}
  \bibinfo{author}{\bibfnamefont{K.}~\bibnamefont{Glampedakis}},
  \bibinfo{journal}{Nucl. Phys.} \textbf{\bibinfo{volume}{A763}},
  \bibinfo{pages}{212} (\bibinfo{year}{2005}), \eprint{astro-ph/0411748}.

\bibitem[{\citenamefont{Harten}(1978)}]{harten1978}
\bibinfo{author}{\bibfnamefont{A.}~\bibnamefont{Harten}},
  \bibinfo{journal}{Mathematics of Computation} \textbf{\bibinfo{volume}{32}},
  \bibinfo{pages}{363} (\bibinfo{year}{1978}),
  \urlprefix\url{http://www.jstor.org/stable/2006149}.

\bibitem[{\citenamefont{Barnes et~al.}(2004)\citenamefont{Barnes, Lefloch,
  Schmidt, and Stewart}}]{Barnes:2004if}
\bibinfo{author}{\bibfnamefont{A.~P.} \bibnamefont{Barnes}},
  \bibinfo{author}{\bibfnamefont{P.~G.} \bibnamefont{Lefloch}},
  \bibinfo{author}{\bibfnamefont{B.~G.} \bibnamefont{Schmidt}},
  \bibnamefont{and} \bibinfo{author}{\bibfnamefont{J.~M.}
  \bibnamefont{Stewart}}, \bibinfo{journal}{Class. Quant. Grav.}
  \textbf{\bibinfo{volume}{21}}, \bibinfo{pages}{5043} (\bibinfo{year}{2004}).

\bibitem[{\citenamefont{Thornburg}(2007)}]{Thornburg:2006zb}
\bibinfo{author}{\bibfnamefont{J.}~\bibnamefont{Thornburg}},
  \bibinfo{journal}{Living Reviews in Relativity} \textbf{\bibinfo{volume}{10}}
  (\bibinfo{year}{2007}),
  \urlprefix\url{http://www.livingreviews.org/lrr-2007-3}.

\bibitem[{\citenamefont{Diener}(2003)}]{Diener:2003jc}
\bibinfo{author}{\bibfnamefont{P.}~\bibnamefont{Diener}},
  \bibinfo{journal}{Class. Quant. Grav.} \textbf{\bibinfo{volume}{20}},
  \bibinfo{pages}{4901} (\bibinfo{year}{2003}), \eprint{gr-qc/0305039}.

\bibitem[{\citenamefont{Osher and Fedkiw}(2003)}]{OsherFedkiw2003}
\bibinfo{author}{\bibfnamefont{S.}~\bibnamefont{Osher}} \bibnamefont{and}
  \bibinfo{author}{\bibfnamefont{R.}~\bibnamefont{Fedkiw}},
  \emph{\bibinfo{title}{Level {S}et {M}ethods and {D}ynamic {I}mplicit
  {S}urfaces}}, vol. \bibinfo{volume}{153} of \emph{\bibinfo{series}{Applied
  Mathematical Sciences}} (\bibinfo{publisher}{{S}pringer-{V}erlag},
  \bibinfo{year}{2003}).

\bibitem[{\citenamefont{Sethian}(1996)}]{Sethian1996}
\bibinfo{author}{\bibfnamefont{J.}~\bibnamefont{Sethian}},
  \emph{\bibinfo{title}{Level {S}et {M}ethods and {F}ast {M}arching {M}ethods}}
  (\bibinfo{publisher}{Cambridge University Press}, \bibinfo{year}{1996}).

\bibitem[{\citenamefont{Fedkiw et~al.}(1999)\citenamefont{Fedkiw, Aslam,
  Merriman, and Osher}}]{Fedkiw1999457}
\bibinfo{author}{\bibfnamefont{R.~P.} \bibnamefont{Fedkiw}},
  \bibinfo{author}{\bibfnamefont{T.}~\bibnamefont{Aslam}},
  \bibinfo{author}{\bibfnamefont{B.}~\bibnamefont{Merriman}}, \bibnamefont{and}
  \bibinfo{author}{\bibfnamefont{S.}~\bibnamefont{Osher}},
  \bibinfo{journal}{Journal of Computational Physics}
  \textbf{\bibinfo{volume}{152}}, \bibinfo{pages}{457 } (\bibinfo{year}{1999}),
  ISSN \bibinfo{issn}{0021-9991},
  \urlprefix\url{http://www.sciencedirect.com/science/article/B6WHY-45GMW83-31%
/2/78188ca59df6a26d48c72b85a6dfcf4e}.

\bibitem[{\citenamefont{Liu et~al.}(2003)\citenamefont{Liu, Khoo, and
  Yeo}}]{Liu2003651}
\bibinfo{author}{\bibfnamefont{T.~G.} \bibnamefont{Liu}},
  \bibinfo{author}{\bibfnamefont{B.~C.} \bibnamefont{Khoo}}, \bibnamefont{and}
  \bibinfo{author}{\bibfnamefont{K.~S.} \bibnamefont{Yeo}},
  \bibinfo{journal}{Journal of Computational Physics}
  \textbf{\bibinfo{volume}{190}}, \bibinfo{pages}{651 } (\bibinfo{year}{2003}),
  ISSN \bibinfo{issn}{0021-9991},
  \urlprefix\url{http://www.sciencedirect.com/science/article/B6WHY-49324BB-8/%
2/09c039d91ac49e11ff1dba4f5d07e88a}.

\bibitem[{\citenamefont{Liu et~al.}(2005)\citenamefont{Liu, Khoo, and
  Wang}}]{Liu2005193}
\bibinfo{author}{\bibfnamefont{T.}~\bibnamefont{Liu}},
  \bibinfo{author}{\bibfnamefont{B.}~\bibnamefont{Khoo}}, \bibnamefont{and}
  \bibinfo{author}{\bibfnamefont{C.}~\bibnamefont{Wang}},
  \bibinfo{journal}{Journal of Computational Physics}
  \textbf{\bibinfo{volume}{204}}, \bibinfo{pages}{193 } (\bibinfo{year}{2005}),
  ISSN \bibinfo{issn}{0021-9991},
  \urlprefix\url{http://www.sciencedirect.com/science/article/B6WHY-4DTKKXN-1/%
2/6155ce6fc7f419d370ba0b8c25d8f5c2}.

\bibitem[{\citenamefont{Banyuls et~al.}(1997)\citenamefont{Banyuls, Font,
  Ibanez, Marti, , and Miralles}}]{Banyuls97}
\bibinfo{author}{\bibfnamefont{F.}~\bibnamefont{Banyuls}},
  \bibinfo{author}{\bibfnamefont{J.~A.} \bibnamefont{Font}},
  \bibinfo{author}{\bibfnamefont{J.~M.} \bibnamefont{Ibanez}},
  \bibinfo{author}{\bibfnamefont{J.~M.} \bibnamefont{Marti}}, ,
  \bibnamefont{and} \bibinfo{author}{\bibfnamefont{J.~A.}
  \bibnamefont{Miralles}}, \bibinfo{journal}{The Astrophysical Journal}
  \textbf{\bibinfo{volume}{476}}, \bibinfo{pages}{221} (\bibinfo{year}{1997}),
  \urlprefix\url{http://stacks.iop.org/0004-637X/476/221}.

\bibitem[{\citenamefont{{Romero} et~al.}(1996)\citenamefont{{Romero}, {Ibanez},
  {Marti}, and {Miralles}}}]{Romero:1995cn}
\bibinfo{author}{\bibfnamefont{J.~V.} \bibnamefont{{Romero}}},
  \bibinfo{author}{\bibfnamefont{J.~M.~A.} \bibnamefont{{Ibanez}}},
  \bibinfo{author}{\bibfnamefont{J.~M.~A.} \bibnamefont{{Marti}}},
  \bibnamefont{and} \bibinfo{author}{\bibfnamefont{J.~A.}
  \bibnamefont{{Miralles}}}, \bibinfo{journal}{\apj}
  \textbf{\bibinfo{volume}{462}}, \bibinfo{pages}{839} (\bibinfo{year}{1996}),
  \eprint{arXiv:astro-ph/9509121}.

\bibitem[{\citenamefont{Neilsen}(1999)}]{NeilsenThesis}
\bibinfo{author}{\bibfnamefont{D.~W.} \bibnamefont{Neilsen}}, Ph.D. thesis,
  \bibinfo{school}{The University of Texas at Austin} (\bibinfo{year}{1999}).

\bibitem[{\citenamefont{Noble}(2003)}]{NobleThesis}
\bibinfo{author}{\bibfnamefont{S.~C.} \bibnamefont{Noble}}, Ph.D. thesis,
  \bibinfo{school}{The University of Texas at Austin} (\bibinfo{year}{2003}),
  \bibinfo{note}{gr-qc/0310116}.

\bibitem[{\citenamefont{Hawke et~al.}(2005)\citenamefont{Hawke, Loffler, and
  Nerozzi}}]{Hawke:2005zw}
\bibinfo{author}{\bibfnamefont{I.}~\bibnamefont{Hawke}},
  \bibinfo{author}{\bibfnamefont{F.}~\bibnamefont{Loffler}}, \bibnamefont{and}
  \bibinfo{author}{\bibfnamefont{A.}~\bibnamefont{Nerozzi}},
  \bibinfo{journal}{Phys. Rev.} \textbf{\bibinfo{volume}{D71}},
  \bibinfo{pages}{104006} (\bibinfo{year}{2005}), \eprint{gr-qc/0501054}.

\bibitem[{\citenamefont{van Leer}(1973)}]{vanLeer73}
\bibinfo{author}{\bibfnamefont{B.~J.} \bibnamefont{van Leer}},
  \bibinfo{journal}{Lecture Notes in Physics} \textbf{\bibinfo{volume}{18}},
  \bibinfo{pages}{163} (\bibinfo{year}{1973}).

\bibitem[{\citenamefont{Colella and Woodward}(1984)}]{Colella1984174}
\bibinfo{author}{\bibfnamefont{P.}~\bibnamefont{Colella}} \bibnamefont{and}
  \bibinfo{author}{\bibfnamefont{P.~R.} \bibnamefont{Woodward}},
  \bibinfo{journal}{Journal of Computational Physics}
  \textbf{\bibinfo{volume}{54}}, \bibinfo{pages}{174 } (\bibinfo{year}{1984}),
  ISSN \bibinfo{issn}{0021-9991},
  \urlprefix\url{http://www.sciencedirect.com/science/article/B6WHY-4DD1PHM-SJ%
/2/13d69a59afba3d6a5d6bbf1144d860aa}.

\bibitem[{\citenamefont{Mart{\'\i} and M{\"u}ller}(1996)}]{Marti19961}
\bibinfo{author}{\bibfnamefont{J.~M.} \bibnamefont{Mart{\'\i}}}
  \bibnamefont{and}
  \bibinfo{author}{\bibfnamefont{E.}~\bibnamefont{M{\"u}ller}},
  \bibinfo{journal}{Journal of Computational Physics}
  \textbf{\bibinfo{volume}{123}}, \bibinfo{pages}{1 } (\bibinfo{year}{1996}),
  ISSN \bibinfo{issn}{0021-9991},
  \urlprefix\url{http://www.sciencedirect.com/science/article/B6WHY-45NJMH5-1D%
/2/98a5fc5dfa9977cc99ee640aa2217732}.

\bibitem[{\citenamefont{Einfeldt}(1988)}]{einfeldt:294}
\bibinfo{author}{\bibfnamefont{B.}~\bibnamefont{Einfeldt}},
  \bibinfo{journal}{SIAM Journal on Numerical Analysis}
  \textbf{\bibinfo{volume}{25}}, \bibinfo{pages}{294} (\bibinfo{year}{1988}),
  \urlprefix\url{http://link.aip.org/link/?SNA/25/294/1}.

\bibitem[{\citenamefont{Donat and Marquina}(1996)}]{Donat199642}
\bibinfo{author}{\bibfnamefont{R.}~\bibnamefont{Donat}} \bibnamefont{and}
  \bibinfo{author}{\bibfnamefont{A.}~\bibnamefont{Marquina}},
  \bibinfo{journal}{Journal of Computational Physics}
  \textbf{\bibinfo{volume}{125}}, \bibinfo{pages}{42 } (\bibinfo{year}{1996}),
  ISSN \bibinfo{issn}{0021-9991},
  \urlprefix\url{http://www.sciencedirect.com/science/article/B6WHY-45MGWCH-62%
/2/b14ec3c8a1e12309465c5d9866d54dec}.

\bibitem[{\citenamefont{Shu}(1999)}]{Shu99}
\bibinfo{author}{\bibfnamefont{C.~W.} \bibnamefont{Shu}}, in
  \emph{\bibinfo{booktitle}{High-{O}rder {M}ethods for {C}omputational
  {P}hysics}}, edited by \bibinfo{editor}{\bibfnamefont{T.~J.}
  \bibnamefont{Barth}} \bibnamefont{and}
  \bibinfo{editor}{\bibfnamefont{H.}~\bibnamefont{Deconinck}}
  (\bibinfo{publisher}{Springer}, \bibinfo{year}{1999}).

\bibitem[{\citenamefont{Wang et~al.}(2004)\citenamefont{Wang, Anderson, Oakley,
  Corradini, and Bonazza}}]{Wang2004528}
\bibinfo{author}{\bibfnamefont{S.-P.} \bibnamefont{Wang}},
  \bibinfo{author}{\bibfnamefont{M.~H.} \bibnamefont{Anderson}},
  \bibinfo{author}{\bibfnamefont{J.~G.} \bibnamefont{Oakley}},
  \bibinfo{author}{\bibfnamefont{M.~L.} \bibnamefont{Corradini}},
  \bibnamefont{and} \bibinfo{author}{\bibfnamefont{R.}~\bibnamefont{Bonazza}},
  \bibinfo{journal}{Journal of Computational Physics}
  \textbf{\bibinfo{volume}{195}}, \bibinfo{pages}{528 } (\bibinfo{year}{2004}),
  ISSN \bibinfo{issn}{0021-9991},
  \urlprefix\url{http://www.sciencedirect.com/science/article/B6WHY-4B3NM1S-2/%
2/96f62d6bfc07ea12490c213d8a1ab263}.

\bibitem[{\citenamefont{Dolezal and Wong}(1995)}]{Dolezal1995266}
\bibinfo{author}{\bibfnamefont{A.}~\bibnamefont{Dolezal}} \bibnamefont{and}
  \bibinfo{author}{\bibfnamefont{S.~S.~M.} \bibnamefont{Wong}},
  \bibinfo{journal}{Journal of Computational Physics}
  \textbf{\bibinfo{volume}{120}}, \bibinfo{pages}{266 } (\bibinfo{year}{1995}),
  ISSN \bibinfo{issn}{0021-9991},
  \urlprefix\url{http://www.sciencedirect.com/science/article/B6WHY-45NJJFW-1D%
/2/8df2ef1b62e40d14d6b0e07fcda05dbd}.

\bibitem[{\citenamefont{{Del Zanna} and {Bucciantini}}(2002)}]{DelZanna02}
\bibinfo{author}{\bibfnamefont{L.}~\bibnamefont{{Del Zanna}}} \bibnamefont{and}
  \bibinfo{author}{\bibfnamefont{N.}~\bibnamefont{{Bucciantini}}},
  \bibinfo{journal}{\aap} \textbf{\bibinfo{volume}{390}}, \bibinfo{pages}{1177}
  (\bibinfo{year}{2002}).

\bibitem[{\citenamefont{Lombard and Donat}(2005)}]{lombard:208}
\bibinfo{author}{\bibfnamefont{B.}~\bibnamefont{Lombard}} \bibnamefont{and}
  \bibinfo{author}{\bibfnamefont{R.}~\bibnamefont{Donat}},
  \bibinfo{journal}{SIAM Journal on Scientific Computing}
  \textbf{\bibinfo{volume}{27}}, \bibinfo{pages}{208} (\bibinfo{year}{2005}),
  \urlprefix\url{http://link.aip.org/link/?SCE/27/208/1}.

\bibitem[{\citenamefont{Abgrall and Karni}(2001)}]{Abgrall2001594}
\bibinfo{author}{\bibfnamefont{R.}~\bibnamefont{Abgrall}} \bibnamefont{and}
  \bibinfo{author}{\bibfnamefont{S.}~\bibnamefont{Karni}},
  \bibinfo{journal}{Journal of Computational Physics}
  \textbf{\bibinfo{volume}{169}}, \bibinfo{pages}{594 } (\bibinfo{year}{2001}),
  ISSN \bibinfo{issn}{0021-9991},
  \urlprefix\url{http://www.sciencedirect.com/science/article/B6WHY-45BC2B2-5C%
/2/5494ddc77cb5e2f7ab2c23fe25353675}.

\end{thebibliography}


\end{document}